%% file: main.tex
\documentclass[sigconf]{acmart}
\usepackage{natbib}
\usepackage{mathtools}
\usepackage{multicol, multirow}
\usepackage{threeparttable}
\usepackage{booktabs}
\usepackage{amsmath}

\usepackage{subfigure}
\usepackage{algorithm}
\usepackage{algpseudocode}
\usepackage{ulem}
\usepackage{caption}
\usepackage{bbding}
\usepackage{enumitem}
\usepackage{dutchcal}

\ccsdesc[500]{Information systems~Recommender systems}
\ccsdesc[300]{Information systems~Online advertising}

\keywords{CTR Prediction, Feature Set, Feature Interaction}

\copyrightyear{2023}
\acmYear{2023}
\setcopyright{acmcopyright}\acmConference[WWW '23]{Proceedings of The ACM Web Conference 2023}{April 30--May 4, 2023}{Austin, TX, USA}
\acmBooktitle{Proceedings of The ACM Web Conference 2023 (WWW '23), April 30--May 4, 2023, Austin, TX, USA}

\begin{document}

\title{Optimizing Feature Set for Click-Through Rate Prediction}

\author{Fuyuan Lyu}
\authornote{Both authors contributed equally to this research.}
\affiliation{
  \institution{McGill University}
  \city{Montreal}
  \country{Canada}
}
\email{fuyuan.lyu@mail.mcgill.ca}

\author{Xing Tang}
\authornotemark[1]
\affiliation{
  \institution{FiT, Tencent}
  \city{Shenzhen}
  \country{China}
}
\email{shawntang@tencent.com}

\author{Dugang Liu}
\authornote{This work was done when working at FiT, Tencent.}
\authornotemark[3]
\affiliation{
  \institution{Guangdong Laboratory of Artificial Intelligence and Digital Economy (SZ)}
  \city{Shenzhen}
  \country{China}
}
\email{dugang.ldg@gmail.com}

\author{Liang Chen}
\affiliation{
  \institution{FiT, Tencent}
  \city{Shenzhen}
  \country{China}
}
\email{leocchen@tencent.com}

\author{Xiuqiang He}
\authornote{Corresponding authors}
\affiliation{
  \institution{FiT, Tencent}
  \city{Shenzhen}
  \country{China}
}
\email{xiuqianghe@tencent.com}

\author{Xue Liu}
\affiliation{
  \institution{McGill University}
  \city{Montreal}
  \country{Canada}
}
\email{xueliu@cs.mcgill.ca}
\renewcommand{\shortauthors}{Fuyuan Lyu et al.}

\input{section/abstract}
\maketitle

\input{section/introduction}
\input{section/method}

\input{section/experiment}
\input{section/related_work}
\input{section/conclusion}

\normalem
\bibliographystyle{ACM-Reference-Format}
\bibliography{main.bib}
\end{document}

%% file: section/abstract.tex
\begin{abstract}
Click-through prediction (CTR) models transform features into latent vectors and enumerate possible feature interactions to improve performance based on the input feature set. Therefore, when selecting an optimal feature set, we should consider the influence of both features and their interaction. However, most previous works focus on either feature field selection or only select feature interaction based on the fixed feature set to produce the feature set. The former restricts search space to the feature field, which is too coarse to determine subtle features. They also do not filter useless feature interactions, leading to higher computation costs and degraded model performance. The latter identifies useful feature interaction from all available features, resulting in many redundant features in the feature set. In this paper, we propose a novel method named OptFS to address these problems. To unify the selection of features and their interaction, we decompose the selection of each feature interaction into the selection of two correlated features. Such a decomposition makes the model end-to-end trainable given various feature interaction operations. By adopting feature-level search space, we set a learnable gate to determine whether each feature should be within the feature set. Because of the large-scale search space, we develop a learning-by-continuation training scheme to learn such gates. Hence, OptFS generates the feature set containing features that improve the final prediction results. Experimentally, we evaluate OptFS on three public datasets, demonstrating OptFS can optimize feature sets which enhance the model performance and further reduce both the storage and computational cost.
\end{abstract}

%% file: section/introduction.tex
\section{Introduction}

\begin{figure}[!htbp]
    \centering
    \includegraphics[width=0.4\textwidth]{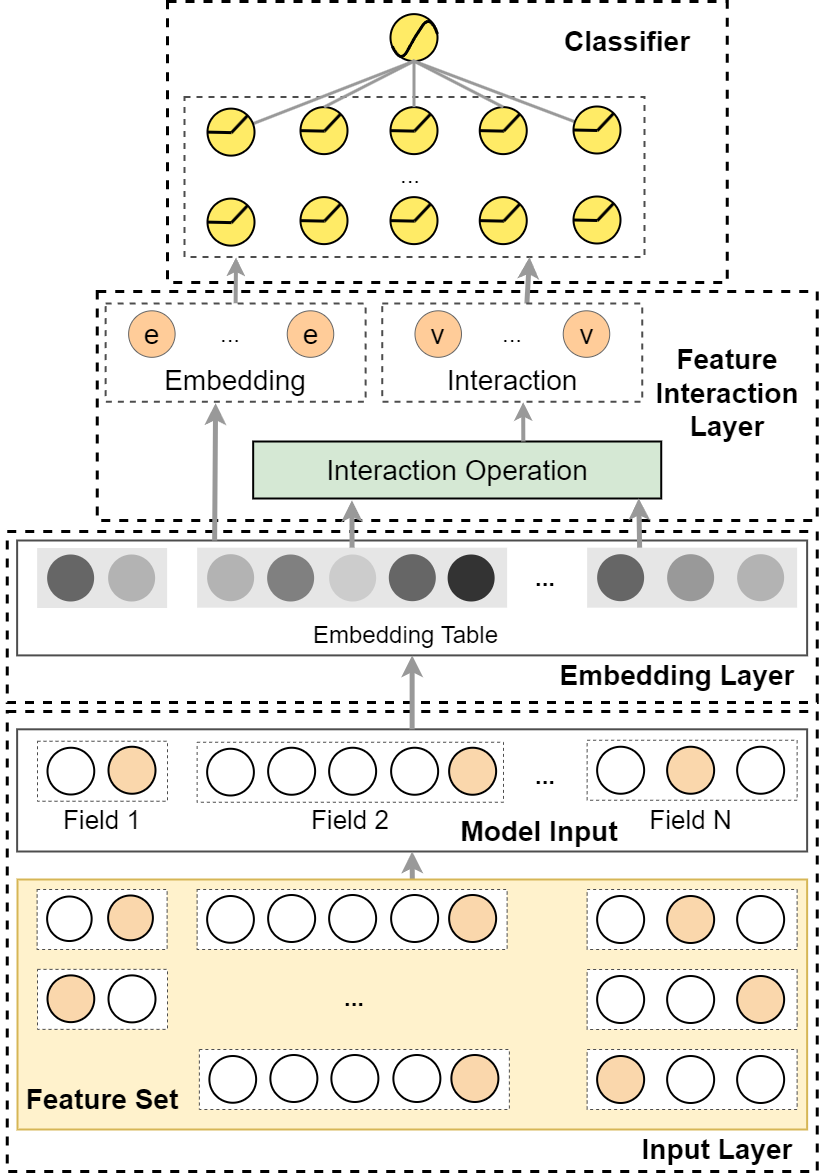}
    \vspace{-10pt}
    \caption{Overview of the general CTR framework.}
    \vspace{-10pt}
    \label{fig:normal}
\end{figure}

Click-through rate prediction has been a crucial task in real-world commercial recommender systems and online advertising systems. It aims to predict the probability of a certain user clicking a recommended item (e.g. movie, advertisement) ~\cite{LR, ADS}. The standard input for CTR prediction consists mainly of a large set of categorical features organized as feature fields. For example, every sample contains a feature field \textit{gender} in CTR prediction, and the field \textit{gender} may include three feature values, \textit{male}, \textit{female} and \textit{unknown}. To avoid ambiguity, we term feature values as features hereafter. General CTR prediction models first map each feature in the feature set into a unique real-valued dense vector through embedding table~\cite{OptEmbed}. Then these vectors are fed into the feature interaction layer to improve the prediction by explicitly modelling low-order feature interaction by enumerating feature set~\cite{AutoPI}. The final prediction of the classifier is made upon the feature embedding and feature interactions, which are both heavily influenced by the input feature set. The general framework is shown in Figure \ref{fig:normal}. Hence, the input feature set plays an important role in CTR prediction.

Blindly inputting all available features into the feature set is neither effective nor efficient. From the view of effectiveness, certain features can be detrimental to model performance. Firstly, these features themselves may only introduce extra learnable parameters, making the prediction model prone to overfitting~\cite{RLReview, Elements_SL}. Secondly, certain useless interactions introduced by these features also bring unnecessary noise and complicate the training process~\cite{AutoFIS}, which degrades the final prediction. Notice that these two factors are closely related when selecting the feature set. If one feature $\mathbf{x}_i$ is filtered out from the set, all its related interactions $\langle \mathbf{x}_i, \cdot \rangle$ should be excluded in the model as well. Correspondingly, informative interactions $\langle \mathbf{x}_i, \mathbf{x}_j \rangle$ is a strong indicator to keep $\mathbf{x}_i$ in the feature set~\cite{AutoCross}. From the view of efficiency, introducing redundant features into a feature set can be inefficient in both storage space and computation cost. As the embedding table dominates the number of parameters in CTR models~\cite{sfctr}, a feature set without redundant features will greatly decrease the size of the models. Moreover, a feature set with useful features can zero out the computation of many useless feature interaction, which greatly reduce the computation cost in practice. An optimal feature set should keep features considering both effectiveness and efficiency.

Efforts have been made to search for an optimal feature set from two aspects. Firstly, Several methods produce the feature set based on feature selection. Because of the large-scale CTR dataset, some methods~\cite{LASSO, LPFS, AutoField} focus on the field level, which results in hundreds of fields instead of millions of features. However, the field level is too coarse to find an optimal feature set. For instance, the feature field \textit{ID} contains user/item feature \textit{id} in real datasets. The \textit{id} of certain cold users/items might be excluded from the feature set due to the sparsity problem~\cite{Cold-start}, which is difficult to handle at the field level. Besides, these methods~\cite{AdaFS, LPFS} fail to leverage the influence of feature interaction, which is commonly considered an enhancement for the model performance~\cite{OptInter,fuxictr}. Secondly, there is also some weakness concerning feature interaction methods, which implicitly produce the feature set. On the one hand, some feature interaction selection methods~\cite{AutoFIS, AutoFeature, OptInter}, inspired by the ideas of neural architecture search~\cite{DARTS, NAO}, tend to work on a fixed subset of input feature set, which commonly includes the redundant features. On the other hand, some method~\cite{AutoCross} constructs a locally optimal feature set to generate feature interaction in separated stages, which requires many handcraft rules to guide the search scheme. Given that many operations of feature interactions are proposed~\cite{DCN, DeepFM, IPNN}, searching an optimal feature set with these operations in a unified way can reduce useless feature interaction. As discussed, optimizing the feature set incorporated with the selection of both feature and feature interaction is required.

In this paper, we propose a method, \textbf{Opt}imizing \textbf{F}eature \textbf{S}et (OptFS), to address the problem of searching the optimal feature set. There are two main challenges for our OptFS. 
The first challenge is how to select the feature and its interaction jointly, given various feature interaction operations. As discussed above, an optimal feature set should exclude features that introduce useless interaction in models. We tackle this challenge by decomposing the selection of each feature interaction into the selection of two correlated features. Therefore, OptFS reduces the search space of feature interaction and trains the model end-to-end, given various feature interaction operations.
The second challenge is the number of features in large-scale datasets. Notice that the possible number of features considered in our research could be $10^6$, which is incredibly larger than $100$ feature fields in previous works~\cite{AutoField, LPFS}. To navigate in the large search space, we introduce a learnable gate for each feature and adopt the learning-by-continuation~\cite{DST, Cont_Spar, Grow_Spar} training scheme. We summarize our major contributions as follows:

\begin{itemize}[topsep=0pt,noitemsep,nolistsep,leftmargin=*]
    \item This paper first distinguishes the optimal feature set problem, which focuses on the feature level and considers the effectiveness of both feature and feature interaction, improving the model performance and computation efficiency.
    \item We propose a novel method named OptFS that optimizes the feature set. Developing an efficient learning-by-continuation training scheme, OptFS leverages feature interaction operations trained together with the prediction model in an end-to-end manner.
    \item Extensive experiments are conducted on three large-scale public datasets. The experimental results demonstrate the effectiveness and efficiency of the proposed method.
\end{itemize}

We organize the rest of the paper as follows. In Section \ref{sec:method}, we formulate the CTR prediction and feature selection problem and propose a simple but effective method OptFS. Section \ref{sec:experiment} details the experiments. In Section \ref{sec:rw}, we briefly introduce related works. Finally, we conclude this paper in Section \ref{sec:conclusion}.

%% file: section/method.tex
\section{OptFS}
\label{sec:method}

In this section, we will first distinguish the feature set optimization problem in Section \ref{sec:form} and detail how OptFS conduct feature selection in Section \ref{sec:method_fs}. Then, we will illustrate how OptFS influences feature interaction selection in Section \ref{sec:method_fis}. Finally, we will illustrate the learning-by-continuation method in Section \ref{sec:method_lbc}.

\subsection{Problem Formulation}
\label{sec:form}

In this subsection, we provide a formulation of the feature set optimization problem. Usually, features that benefit the accurate prediction are considered useful in CTR models. In our setting, we represent all possible features as $\mathbf{X} = \{\mathbf{x}_{1}, \mathbf{x}_{2}, \cdots, \mathbf{x}_{m}\}$. $\mathbf{x}_{i}$ is a one-hot representation, which is very sparse and high-dimensional. As previously discussed, the feature set optimization problem aims to determine the useful features among all possible ones, which can be defined as finding an optimal feature set $\mathbf{X}^\mathbf{g} \subset \mathbf{X}$. This can be formulated as follows:

\begin{equation}
\begin{aligned}
    \min_{\mathbf{W}} \mathcal{L}(\mathcal{D}|\mathbf{W}) & , \ 
    \mathcal{D} = \{ \mathbf{X}^\mathbf{g}, \mathbf{Y}\}, \\
    s.t. \forall \mathbf{x} \in \mathbf{X}^\mathbf{g}, \mathcal{L}(\mathbf{X}^\mathbf{g}) & > \mathcal{L}(\mathbf{X}^\mathbf{g}-\{\mathbf{x}\}), \\
    \forall \mathbf{x} \notin \mathbf{X}^\mathbf{g}, \mathcal{L}(\mathbf{X}^\mathbf{g}) & \ge \mathcal{L}(\mathbf{X}^\mathbf{g}+\{\mathbf{x}\}),
\end{aligned}
\end{equation}
where $\mathcal{L}$ denotes the loss function, $\mathbf{W}$ denotes the model parameters, and $\mathbf{Y}$ denotes the corresponding labels.

\subsection{Feature Selection}
\label{sec:method_fs}
Each field $\mathbf{z}_i$ contains a proportion of all possible features, denoted as:
\begin{equation}
\label{eq:set}
    \mathbf{z}_i = \{ \mathbf{x}_{k_i} \}, \ 1 \le k_i \le m,
\end{equation}
which indicates that the relationship between field and feature is a one-to-many mapping. In practice, the number of field $n$ is much smaller than that of feature $m$. For instance, online advertisement systems usually have $n \le 100$ and $m \approx 10^6$. So the input of CTR models can be rewritten as follows from both feature and field perspectives:
\begin{equation}
\label{eq:input}
    \mathbf{z} = [\mathbf{z}_1, \mathbf{z}_2, \cdots, \mathbf{z}_n] = [\mathbf{x}_{k_1}, \mathbf{x}_{k_2}, \cdots, \mathbf{x}_{k_n}],
\end{equation}
where the second equal sign means that for input $\mathbf{z}$, the corresponding feature for field $\mathbf{z}_i$ is $\mathbf{x}_{k_i}$ as shown in Equation \ref{eq:set}. 

We usually employ embedding tables to convert $\mathbf{z}_i$s into low-dimensional and dense real-value vectors. This can be formulated as $\mathbf{e}_{i}=\mathbf{E} \times \mathbf{z}_{i}=\mathbf{E} \times \mathbf{x}_{k_i}, 1 \le i \le n, 1 \le k_i \le m$, where $\mathbf{E}\in\mathbb{R}^{m\times D}$ is the embedding table, $m$ is the number of feature values and $D$ is the size of embedding. Then embeddings are stacked together as a embedding vector $\mathbf{e} = [\mathbf{e}_{1}, \mathbf{e}_{2}, \cdots, \mathbf{e}_{n}]$.

In our work, we propose feature-level selection. Instead of doing field-level selection, we formulate selection as assigning a binary gate $\mathbf{g}_{k_i}\in\{0,1\}$ for each feature embedding $\mathbf{e}_{k_i}$. After selection, the feature embeddings can be formulated as follows: 
\begin{equation}
    \mathbf{e}_{k_i}^\mathbf{g} = \mathbf{g}_{k_i} \odot \mathbf{e}_{k_i} = \mathbf{g}_{k_i} \odot (\mathbf{E} \times \mathbf{x}_{k_i}).
\end{equation}
When $\mathbf{g}_{k_i} = 1$, feature $\mathbf{x}_{k_i}$ is in the optimal feature set $\mathbf{X}^\mathbf{g}$ and vice versa. Notice that previous work~\cite{LASSO, LPFS, AutoField} assigns field-level feature selection. This means that $ \mathbf{g}_{k_i} \equiv \mathbf{g}_{i}\in\{0,1\}$ for each field $z_i$, indicating the keep or drop of all possible features $\{ \mathbf{x}_{k_i} \}$ in corresponding field. 

Then, these embeddings are stacked together as a feature-selected embedding vector $\mathbf{e}^\mathbf{g} = [\mathbf{e}_{k_1}^\mathbf{g}, \mathbf{e}_{k_2}^\mathbf{g}, \cdots, \mathbf{e}_{k_n}^\mathbf{g}]$. The final prediction can be formulated as follows:
\begin{equation}
\label{eq:fs}
    \hat{y} = \mathcal{F}(\mathbf{g} \odot \mathbf{E} \times \mathbf{x}|\mathbf{W}) = \mathcal{F}(\mathbf{E}^{\mathbf{g}} \times \mathbf{x}|\mathbf{W}),
\end{equation}
where $\mathbf{g}\in\{0,1\}^{m}$ refers to gating vectors indicating whether certain feature is selected or not, $\mathbf{E}^\mathbf{g} = \mathbf{g} \odot \mathbf{E}$ indicates the feature-selected embedding tables. The $\mathbf{E}^\mathbf{g}$ can also be viewed as the feature set $\mathbf{X}^\mathbf{g}$ after transformation from the embedding table, denoted as $\mathbf{E}^\mathbf{g} = \mathbf{E} \times \mathbf{X}^\mathbf{g}$.

\subsection{Feature Interaction Selection}
\label{sec:method_fis}

The feature interaction selection aims to select beneficial feature interaction for explicitly modelling~\cite{OptInter, AutoFIS}. The feature interaction layer will be performed based on $\mathbf{e}$ in mainstream CTR models. There are several types of feature interaction in previous study~\cite{AutoFeature}, e.g. inner product~\cite{DeepFM}. The interaction between two features $\mathbf{e}_i$ and $\mathbf{e}_j$ can be generally represented as:
\begin{equation}
    \mathbf{v}_{(i,j)} = \mathcal{O}(\mathbf{e}_i, \mathbf{e}_j),
\end{equation}
where $\mathcal{O}$, as the interaction operation, can vary from a single layer perceptron to cross layer\cite{DCN}. The feature interaction selection can be formulated as assigning $\mathbf{g}^{'}_{(i,j)}$ for each feature interaction. All feature interactions can be aggregated together for final prediction:
\begin{equation}
    \hat{y} = \mathcal{H}((\mathbf{g}^{'} \odot \mathbf{v}) \oplus \mathcal{G}(\mathbf{e}^\mathbf{g})) = \mathcal{H}(\mathbf{v}^{\mathbf{g}^{'}} \oplus \mathcal{G}(\mathbf{e}^\mathbf{g})),
    \label{eq:aggregation}
\end{equation}
where symbol $\oplus$ denotes the concatenation operation, $\mathcal{G}(\cdot)$ denotes the transformation function from embedding space to feature interaction space, such as MLP~\cite{DeepFM,DCN} or null function~\cite{IPNN}. $\mathcal{H}(\cdot)$ represents the prediction function. The combinations of $\mathcal{G}(\cdot)$, $\mathcal{O}(\cdot)$ and $\mathcal{H}(\cdot)$ in mainstream models are summarized in Table \ref{Table:summary}.

\begin{table}[!htbp]
    \centering
    \caption{Summary of $\mathcal{G}(\cdot)$, $\mathcal{O}(\cdot)$ and $\mathcal{H}(\cdot)$ in mainstream models}
    \begin{tabular}{c|c|c|c}
    \hline
        Model & $\mathcal{G}(\cdot)$ & $\mathcal{O}(\cdot)$ & $\mathcal{H}(\cdot)$ \\
    \hline
        FM~\cite{FM}            & null & inner product & null \\
        DeepFM~\cite{DeepFM}    & MLP & inner product & average \\ 
        DCN~\cite{DCN}          & MLP & cross network & average \\
        IPNN~\cite{IPNN}        & null & inner product & MLP \\
        OPNN~\cite{IPNN}        & null & outer product & MLP \\
        PIN~\cite{PIN}          & null & MLP & MLP \\
    \hline
    \end{tabular}
    \label{Table:summary}
\end{table}

\begin{figure}[!t]
    \centering
    \includegraphics[width=0.45\textwidth]{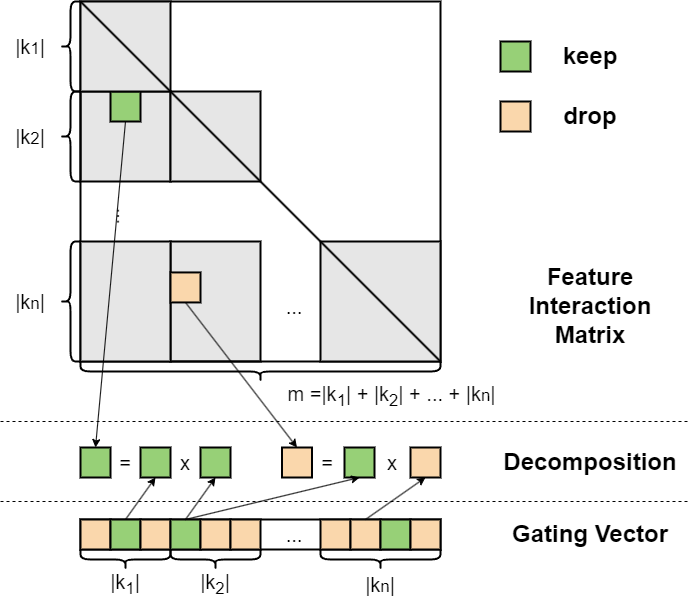}
    \vspace{-10pt}
    \caption{The Overview of OptFS.}
    \vspace{-10pt}
    \label{fig:overall}
\end{figure}

In reality, a direct way to explore all possible feature interaction is introducing a feature interaction matrix $\{\mathbf{g}^{'}_{(k_i,k_j)}\}$ for 2nd-order feature interaction $\{\mathbf{x}_{k_i}, \mathbf{x}_{k_j}\}$. But it is impossible as we would have $C_m^2 \approx 10^{12}$ gate variables. To efficiently narrow down such a large space, previous works~\cite{AutoFIS, AutoFeature, OptInter} restrict the search space to feature field interaction, reducing the number of variables to $C_n^2 \approx 1000$. This can be formulated as $\mathbf{g}^{'}_{(i,j)} \equiv \mathbf{g}^{'}_{(k_i,k_j)}$. However, such relaxation may not be able to distinguish the difference between useful and useless feature interaction within the same field. 
As it has been proven that informative interaction between features tends to come from the informative lower-order ones~\cite{FIVES}, we decompose the feature interaction as follows:
\begin{equation}
    \mathbf{g}^{'}_{(k_i,k_j)} = \mathbf{g}_{k_i} \times \mathbf{g}_{k_j},
\end{equation}
which indicates that the feature interaction is only deemed useful when both features are useful. An illustration of the decomposition is shown in Figure \ref{fig:overall}. Hence, the final prediction can be written as:
\begin{equation}
\label{eq:final_predict}
    \hat{y} = \mathcal{H}((\mathbf{g} \times \mathbf{g} \odot \mathbf{v}) \oplus \mathcal{G}(\mathbf{g} \odot \mathbf{e})),
\end{equation}
which means that the gating vector $\mathbf{g}$ that selects features can also select the feature interaction given $\mathcal{O}(\cdot)$. Such a design can reduce the search space and obtain the optimal feature set in an end-to-end manner.

\subsection{Learning by Continuation}
\label{sec:method_lbc}
Even though the search space has been narrowed down from $C^2_m + m$ to $m$ in Section \ref{sec:method_fis}, we still need to determine whether to keep or drop each feature in the feature set. This can be formulated as a $\mathcal{l}_0$ normalization problem. However, binary gate vector $m$ is hard to compute valid gradient. Moreover, $\mathcal{l}_0$ optimization is known as a NP-hard problem~\cite{l2norm}. To efficiently train the entire model, we introduce a learning-by-continuation training scheme. Such a training scheme has proven to be an efficient method for approximating $\mathcal{l}_0$ normalization~\cite{Cont_Spar}, which correlates with our goal.

The learning-by-continuation training scheme consists of two parts: the searching stage that determines the gating vector $\mathbf{g}$ and the rewinding stage that determines the embedding table $\mathbf{e}$ and other parameters $\mathbf{W}$. We will introduce them separately in the following sections.

\subsubsection{Searching}
\begin{figure}[!htbp]
    \centering
    \subfigure[Searching Stage]{
    \begin{minipage}[t]{0.22\textwidth}
    \centering
    \includegraphics[width=\textwidth]{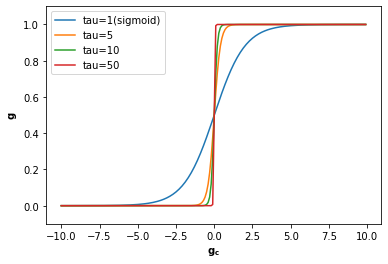}
    \vspace{-5pt}
    \label{fig:tau-training}
    \vspace{-5pt}
    \end{minipage}
    }
    \subfigure[Re-training Stage]{
    \begin{minipage}[t]{0.22\textwidth}
    \centering
    \includegraphics[width=\textwidth]{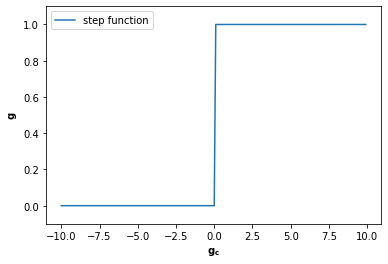}
    \vspace{-5pt}
    \label{fig:tau-inference}
    \vspace{-5pt}
    \end{minipage}
    }
    \vspace{-10pt}
    \caption{Visualization of gating vector $g$ during searching and retraining stages.}
    \vspace{-5pt}
    \label{fig:tau}
\end{figure}

To efficiently optimize the feature set with feature-level granularity, we introduce a continual gate $\mathbf{g}_c\in\mathbb{R}^{m}$. During the searching stage, we introduce an exponentially-increased temperature value $\tau$ to approximate $L_0$ normalization. Specifically, the actual gate $\mathbf{g}$ is computed as:
\begin{equation}
\label{eq:tau}
    \mathbf{g} = \frac{\sigma(\mathbf{g}_c \times \tau)}{\sigma(\mathbf{g}_c^{(0)})}, \ \tau = \gamma^{t/T}
\end{equation}
where $\mathbf{g}_c^{(0)}$ is the initial value of the continual gate $\mathbf{g}_c$, $\sigma$ is the sigmoid function $\sigma(x) = \frac{1}{1+e^{-x}}$ applied element-wise, $t$ is the current training epoch number, $T$ is the total training epoch and $\gamma$ is the final value of $\tau$ after training for $T$ epochs. This would allow the continuous gating vector $\mathbf{g}_c$ to receive valid gradients in early stages yet increasingly approximate binary gate as the epoch number $t$ grows. An illustration of Equation \ref{eq:tau} is shown in Figure \ref{fig:tau-training}.

The final prediction is calculated based on Equation \ref{eq:final_predict}. The cross-entropy loss (i.e. log-loss) is adopted for each sample:
\begin{equation}
\label{eq:logloss}
    \text{CE} (y,\hat{y}) = y\log(\hat{y}) + (1-y)\log(1-\hat{y}),
\end{equation}
where $y$ is the ground truth of user clicks. We summarize the final accuracy loss as follows:
\begin{equation}
\label{eq:summarize}
    \mathcal{L}_{\text{CE}}(\mathcal{D}|\{\mathbf{E}, \mathbf{W}\}) = -\frac{1}{ |\mathcal{D}|} \sum_{(\mathbf{x}, y)\in\mathcal{D}} \text{CE}(y, \mathcal{F}(\mathbf{E} \times \mathbf{x}|\mathbf{W})),
\end{equation}
where $\mathcal{D}$ is the training dataset and $\mathbf{W}$ is network parameters except the embedding table $\mathbf{E}$. Hence, the final training objective becomes:
\begin{equation}
\label{eq:loss_search}
    \min_{\mathbf{g}_c, \mathbf{E}, \mathbf{W}} \ \mathcal{L}_{\text{CE}}(\mathcal{D}|\{\mathbf{g}_c \odot \mathbf{E}, \mathbf{W}\}) + \lambda \lVert \mathbf{g} \rVert_1,
\end{equation}
where $\lambda$ is the regularization penalty, $\lVert \cdot \rVert_1$ indicates the $\mathcal{l}_1$ norm to encourage sparsity. Here we restate $\mathcal{l}_0$ norm to $\mathcal{l}_1$ norm given the fact that $\lVert \mathbf{g} \rVert_0 = \lVert \mathbf{g}\rVert_1$ for binary $\mathbf{g}$.

After training $T$ epochs, the final gating vector $\mathbf{g}$ is calculated through a unit-step function as follows:
\begin{equation}
\label{eq:discretization}
    \mathbf{g} = 
\begin{cases}
    0,\quad &\mathbf{g}_c \leq 0 \\
    1,\quad &\text{otherwise}
\end{cases}.
\end{equation}
Such a unit step function is also visualized in Figure \ref{fig:tau-inference}.

\subsubsection{Retraining}
\label{sec:method_lbc_retrain}
In the searching stage, all possible features are fed into the model to explore the optimal feature set $\mathbf{X}^\mathbf{g}$. Thus, the useless features might hurt the model's performance. To address this problem, we need to retrain the model after obtaining the optimal feature set $\mathbf{X}^\mathbf{g}$. 

After determining the gating vector $\mathbf{g}$, we retrain the model parameters $\mathbf{E}$ and $\mathbf{W}$ as the corresponding values at $T_c$ epoch, which is carefully tuned in our setting. This is because most CTR models early stopped in several epochs, making them more sensitive towards initialization and prone to overfitting~\cite{CTR-Overfit}. The final parameters $\mathbf{E}$ and $\mathbf{W}$ are trained as follows:
\begin{equation}
\label{eq:loss_retrain}
    \min_{\mathbf{E}, \mathbf{W}} \ \mathcal{L}_{\text{CE}}(\mathcal{D}|\{\mathbf{g} \odot \mathbf{E}, \mathbf{W}\}).
\end{equation}
The overall process of our model is summarized in Algorithm \ref{alg:optfs}.

\begin{algorithm}
	\caption{The OptFS Algorithm} 
    \label{alg:optfs}
	\begin{algorithmic}[1]
		\Require training dataset $\mathcal{D}$, initialization epoch $T_c$, total epoch $T$
        \Ensure gating vector $\mathbf{g}$, model parameters \{$\mathbf{E}$, $\mathbf{W}$\}
        
        \State \textbf{\#\# Searching \#\#}
        
        \State t=0
        \While {t < T}
            \State t = t + 1
            \While{$\mathcal{D}$ is not fully iterated}
                \State Sample a mini-batch from the training dataset
                \State$\{\mathbf{E}_t,\mathbf{W}_t\}$, $\mathbf{g}$ = Searching($\mathcal{D}$) \Comment{Equation \ref{eq:loss_search}}
            \EndWhile
            \If{$t == T_c$}
            \State $\{\hat{\mathbf{E}}, \hat{\mathbf{W}}\} \Leftarrow \{\mathbf{E}_t,\mathbf{W}_t\}$
            \EndIf
        \EndWhile 
        \State $\mathbf{g}$ = Discretization($\{\mathbf{g}_c\}$) \Comment{Equation \ref{eq:discretization}}

        \State \textbf{\#\# Retraining\#\#}

        \State Retrain $\{\mathbf{E},\mathbf{W}\}$ given $\mathbf{g}$ with $\{\hat{\mathbf{E}}, \hat{\mathbf{W}}\}$ as initialization \Comment{Equation \ref{eq:loss_retrain}} 
	\end{algorithmic}
\end{algorithm}

%% file: section/experiment.tex
\section{experiment}
\label{sec:experiment}

In this section, to comprehensively evaluate our proposed method, we design experiments to answer the following research questions: 

\begin{itemize}[topsep=0pt,noitemsep,nolistsep,leftmargin=*]
    \item \textbf{RQ1}: Could OptFS achieve superior performance compared with mainstream feature (interaction) selection methods?
    \item \textbf{RQ2}: How does the end-to-end training scheme influence the model performance?
    \item \textbf{RQ3}: How does the re-training stage influence the performance?
    \item \textbf{RQ4}: How efficient is OptFS compared to other feature (interaction) selection methods?
    \item \textbf{RQ5}: Does OptFS select the optimal features?
\end{itemize}

\begin{table*}[!htbp]
\centering
\caption{Performance Comparison Between OptFS and Feature Selection Methods.}	\label{Table:FS}
\vspace{-5pt}
 \resizebox{1.0\textwidth}{!}{
\begin{tabular}{c|c|ccc|ccc|ccc|ccc}
    \hline
        & \multirow{2}{*}{Method} & \multicolumn{3}{c}{FM} & \multicolumn{3}{|c}{DeepFM} & \multicolumn{3}{|c}{DCN} & \multicolumn{3}{|c}{IPNN} \\
    \cline{3-14}
        &  & AUC$\uparrow$ & Logloss$\downarrow$ & Ratio$\downarrow$ & AUC$\uparrow$ & Logloss$\downarrow$ & Ratio$\downarrow$ & AUC$\uparrow$ & Logloss$\downarrow$ & Ratio$\downarrow$ & AUC$\uparrow$ & Logloss$\downarrow$ & Ratio$\downarrow$ \\
    \hline
        \multirow{5}{*}{\rotatebox{90}{Criteo}}
            & Backbone  & 0.8055 & 0.4457 & 1.0000 & 0.8089 & 0.4426 & 1.0000 & 0.8107 & 0.4410 & 1.0000 & 0.8110 & 0.4407 & 1.0000 \\
            & LPFS      & 0.7888 & 0.4604 & 0.0157 & 0.7915 & 0.4579 & 0.2415 & 0.7802 & 0.4743 & 0.1177 & 0.7789 & 0.4705 & 0.3457 \\
            & AutoField & 0.7932 & 0.4567 & \textbf{0.0008} & 0.8072 & 0.4439 & 0.3811 & \textbf{0.8113} & \textbf{0.4402} & 0.5900 & 0.8115 & \textbf{0.4401} & 0.9997 \\
            & AdaFS     & 0.7897 & 0.4597 & 1.0000 & 0.8005 & 0.4501 & 1.0000 & 0.8053 & 0.4472 & 1.0000 & 0.8065 & 0.4448 & 1.0000 \\
            & OptFS     & \textbf{0.8060} & \textbf{0.4454} & 0.1387 & $\textbf{0.8100}^*$ & $\textbf{0.4415}^*$ & \textbf{0.0422} & 0.8111 & 0.4405 & \textbf{0.0802} & \textbf{0.8116} & \textbf{0.4401} & \textbf{0.0719} \\
    \hline
        \multirow{5}{*}{\rotatebox{90}{Avazu}}
            & Backbone  & 0.7838 & 0.3788 & 1.0000 & 0.7901 & 0.3757 & 1.0000 & 0.7899 & 0.3755 & 1.0000 & 0.7913 & 0.3744 & 1.0000 \\
            & LPFS      & 0.7408 & 0.4029 & 0.7735 & 0.7635 & 0.3942 & 0.9975 & 0.7675 & 0.3889 & 0.9967 & 0.7685 & 0.3883 & 0.9967 \\
            & AutoField & 0.7680 & 0.3862 & \textbf{0.0061} & 0.7870 & 0.3773 & 1.0000 & 0.7836 & 0.3782 & 0.9992 & 0.7865 & 0.3770 & 0.9992 \\
            & AdaFS     & 0.7596 & 0.3913 & 1.0000 & 0.7797 & 0.3837 & 1.0000 & 0.7693 & 0.3954 & 1.0000 & 0.7818 & 0.3833 & 1.0000 \\
            & OptFS     & $\textbf{0.7839}$ & $\textbf{0.3784}$ & 0.8096 & $\textbf{0.7946}^*$ & $\textbf{0.3712}^*$ & \textbf{0.8686} & $\textbf{0.7932}^*$ & $\textbf{0.3718}^*$ & \textbf{0.8665} & $\textbf{0.7950}^*$ & $\textbf{0.3709}^*$ & \textbf{0.9118} \\
    \hline
        \multirow{5}{*}{\rotatebox{90}{KDD12}}
            & Backbone  & 0.7783 & 0.1566 & 1.0000 & 0.7967 & 0.1531 & 1.0000 & 0.7974 & 0.1531 & 1.0000 & 0.7966 & 0.1532 & 1.0000 \\
            & LPFS      & 0.7725 & 0.1578 & 1.0000 & 0.7964 & 0.1532 & 1.0000 & 0.7970 & \textbf{0.1530} & 1.0000 & 0.7967 & 0.1532 & 1.0000 \\
            & AutoField & 0.7411 & 0.1634 & \textbf{0.0040} & 0.7919 & 0.1542 & 0.9962 & 0.7943 & 0.1536 & \textbf{0.8249} & 0.7926 & 0.1541 & \textbf{0.8761} \\
            & AdaFS     & 0.7418 & 0.1644 & 1.0000 & 0.7917 & 0.1543 & 1.0000 & 0.7939 & 0.1538 & 1.0000 & 0.7936 & 0.1539 & 1.0000 \\
            & OptFS     & $\textbf{0.7811}^*$ & $\textbf{0.1560}^*$ & 0.5773 & $\textbf{0.7988}^*$ & $\textbf{0.1527}^*$ & \textbf{0.9046} & $\textbf{0.7987}^*$ & \textbf{0.1527} & 0.8945 & \textbf{0.7976} & \textbf{0.1530} & 0.8762 \\
    \hline
\end{tabular}
}
\begin{tablenotes}
\footnotesize
\item[1] Here $*$ denotes statistically significant improvement (measured by a two-sided t-test with p-value $<0.05$) over the best baseline. \textbf{Bold} font indicates the best-performed method.
\vspace{-10pt}
\end{tablenotes}
\end{table*}

\subsection{Experiment Setup}
\subsubsection{Datasets}
We conduct our experiments on three public real-world datasets. We describe all datasets and the pre-processing steps below.

\textbf{Criteo}\footnote{https://www.kaggle.com/c/criteo-display-ad-challenge} dataset consists of ad click data over a week. It consists of 26 categorical feature fields and 13 numerical feature fields. Following the best practice~\cite{fuxictr}, we discretize each numeric value $x$ to $\lfloor\log^2(x)\rfloor$, if $x>2$; $x=1$ otherwise. We replace infrequent categorical features with a default "OOV" (i.e. out-of-vocabulary) token, with \textit{min\_count}=2.

\textbf{Avazu}\footnote{http://www.kaggle.com/c/avazu-ctr-prediction} dataset contains 10 days of click logs. It has 24 fields with categorical features. Following the best practice~\cite{fuxictr}, we remove the \textit{instance\_id} field and transform the \textit{timestamp} field into three new fields: \textit{hour}, \textit{weekday} and \textit{is\_weekend}. We replace infrequent categorical features with the "OOV" token, with \textit{min\_count}=2.

\textbf{KDD12}\footnote{http://www.kddcup2012.org/c/kddcup2012-track2/data} dataset contains training instances derived from search session logs. It has 11 categorical fields, and the click field is the number of times the user clicks the ad. We replace infrequent features with an "OOV" token, with \textit{min\_count}=2.

\subsubsection{Metrics}
Following the previous works~\cite{FM,DeepFM}, we use the common evaluation metrics for CTR prediction: \textbf{AUC} (Area Under ROC) and \textbf{Log loss} (cross-entropy). Note that $\mathbf{0.1 \%}$ improvement in AUC is considered significant~\cite{IPNN, DeepFM}. To measure the size of the feature set, we normalize it based on the following equation:
\begin{equation}
\label{eq:ratio}
    \text{Ratio} = \text{\#Remaining Features}/m.
\end{equation}

\subsubsection{Baseline Methods and Backbone Models}
\label{sec:experiment_baseline}
We compare the proposed method OptFS with the following feature selection methods: (i) AutoField~\cite{AutoField}: This baseline utilizes neural architecture search techniques~\cite{DARTS} to select the informative features on a field level; (ii) LPFS~\cite{LPFS}: This baseline designs a customized, smoothed-$\mathcal{l}_0$-liked function to select informative fields on a field level; (iii) AdaFS~\cite{AdaFS}: This baseline that selects the most relevant features for each sample via a novel controller network. We apply the above baselines over the following mainstream backbone models: FM~\cite{FM}, DeepFM~\cite{DeepFM}, DCN~\cite{DCN} and IPNN~\cite{IPNN}.

We also compare the proposed method OptFS with a feature interaction selection method: AutoFIS~\cite{AutoFIS}. This baseline utilizes GRDA optimizer to abandon unimportant feature interaction in a field-level manner. We apply AutoFIS over the following backbone models: FM~\cite{FM}, DeepFM~\cite{DeepFM}. We only compare with AutoFIS on FM and DeepFM backbone models because the original paper only provides the optimal hyper-parameter settings and releases source code under these settings.

\subsubsection{Implementation Details}
In this section, we provide the implementation details. For OptFS, (i) General hyper-params: We set the embedding dimension as 16 and batch size as 4096. For the MLP layer, we use three fully-connected layers of size [1024, 512, 256]. Following previous work~\cite{IPNN}, Adam optimizer, Batch Normalization~\cite{BatchNorm} and Xavier initialization~\cite{Xavier} are adopted. We select the optimal learning ratio from \{1e-3, 3e-4, 1e-4, 3e-5, 1e-5\} and $l_2$ regularization from \{1e-3, 3e-4, 1e-4, 3e-5, 1e-5, 3e-6, 1e-6\}. (ii) OptFS hyper-params: we select the optimal regularization penalty $\lambda$ from \{1e-8, 5e-9, 2e-9, 1e-9\}, training epoch $T$ from \{5, 10, 15\}, final value $\gamma$ from \{2e+2, 5e+2, 1e+3, 2e+3, 5e+3, 1e+4\}. During the re-training phase, we reuse the optimal learning ratio and $l_2$ regularization and choose the rewinding epoch $T_c$ from $\{1, 2, \cdots, T-1\}$.
For AutoField and AdaFS, we select the optimal hyper-parameter from the same hyper-parameter domain of OptFS, given the original paper does not provide the hyper-parameter settings. For LPFS and AutoFIS, we reuse the optimal hyper-parameter mentioned in original papers.

Our implementation\footnote{https://github.com/fuyuanlyu/OptFS} is based on a public Pytorch library for CTR prediction\footnote{https://github.com/rixwew/pytorch-fm}. For other baseline methods, we reuse the official implementation for the AutoFIS\footnote{https://github.com/zhuchenxv/AutoFIS}~\cite{AutoFIS} method. Due to the lack of available implementations for the LPFS~\cite{LPFS}, AdaFS\cite{AdaFS} and AutoField\cite{AutoField} methods, we re-implement them based on the details provided by the authors and open-source them to benefit future researchers\footnote{https://github.com/fuyuanlyu/AutoFS-in-CTR}.

\subsection{Overall Performance(RQ1)}

In this section, we conduct two studies to separately compare feature selection methods and feature interaction selection methods in Section \ref{sec:experiment_fs} and \ref{sec:experiment_fis}. Notes that both these methods can be viewed as a solution to the feature set optimization problem.

\subsubsection{Feature Selection}
\label{sec:experiment_fs}
The overall performance of our OptFS and other feature selection baseline methods on four different backbone models using three benchmark datasets are reported in Table \ref{Table:FS}. We summarize our observation below.

Firstly, our OptFS is effective and efficient compared with other baseline methods. OptFS can achieve higher AUC with a lower feature ratio. However, the benefit brought by OptFS differs on various datasets. On Criteo, OptFS tends to reduce the size of the feature set. OptFS can reduce 86\% to 96\% features with improvement not considered significant statistically. On the Avazu and KDD12 datasets, the benefit tends to be both performance boosting and feature reduction. OptFS can significantly increase the AUC by 0.01\% to 0.45\% compared with the backbone model while using roughly 10\% of the features. Note that the improved performance is because OptFS considers feature interaction's influence during selection.
Meanwhile, other feature selection baselines tend to bring performance degradation. This is likely because they adopt the feature field selection. Such a design will inevitably drop useful features or keep useless ones.

Secondly, different datasets behave differently regarding the redundancy of features. For example, on the Criteo dataset, all methods produce low feature ratios, indicating that this dataset contains many redundant features. On the other hand, on the Avazu and KDD12 datasets, all methods produce high feature ratios, suggesting that these two datasets have lower redundancy. 
OptFS can better balance the trade-off between model performance and efficiency compared with other baselines in all datasets.


Finally, field-level feature selection methods achieve different results on various backbone models. Compared to other deep models, FM solely relies on the explicit interaction, i.e. inner product. If one field $\mathbf{z}_i$ is zeroed out during the process, all its related interactions will be zero. The other fields are also lured into zero, as their interaction with field $\mathbf{z}_i$ does not bring any information into the final prediction. Therefore, it can be observed that LPFS has a low feature ratio on Criteo and high feature ratios on Avazu and KDD12 datasets. On the other hand, AutoField generates low feature ratios ($\sim$0\%) on all three datasets. These observations further highlight the necessity of introducing feature-level granularity into the feature set optimization problem as OptFS does. 

\subsubsection{Feature Interaction Selection}
\label{sec:experiment_fis}

\begin{table}[!htbp]
\centering
\caption{Performance Comparison Between OptFS and Feature Interaction Selection Method.}	\label{Table:FIS}
\vspace{-5pt}
\resizebox{0.40\textwidth}{!}{
\begin{tabular}{c|c|c|ccc}
    \hline
        & \multirow{2}{*}{Model} & \multirow{2}{*}{Method} & \multicolumn{3}{c}{Metrics} \\
    \cline{4-6}
        & & & AUC$\uparrow$ & Logloss$\downarrow$ & Ratio$\downarrow$ \\
    \hline
        \multirow{6}{*}{\rotatebox{90}{Criteo}} & \multirow{3}{*}{FM} 
        & Backbone  & 0.8055 & 0.4457 & 1.0000 \\
        & & AutoFIS & \textbf{0.8063} & \textbf{0.4449} & 1.0000 \\
        & & OptFS   & 0.8060 & 0.4454 & \textbf{0.1387} \\
    \cline{2-6}
        & \multirow{3}{*}{DeepFM} 
        & Backbone  & 0.8089 & 0.4426 & 1.0000 \\
        & & AutoFIS & 0.8097 & 0.4418 & 1.0000 \\
        & & OptFS   & \textbf{0.8100} & \textbf{0.4415} & \textbf{0.0422} \\ 
    \hline
        \multirow{6}{*}{\rotatebox{90}{Avazu}} & \multirow{3}{*}{FM} 
        & Backbone  & 0.7838 & 0.3788 & 1.0000 \\
        & & AutoFIS & $\textbf{0.7843}$ & 0.3785 & 1.0000 \\
        & & OptFS   & 0.7839 & $\textbf{0.3784}$ & \textbf{0.8096} \\
    \cline{2-6}
        & \multirow{3}{*}{DeepFM} 
        & Backbone  & 0.7901 & 0.3757 & 1.0000 \\
        & & AutoFIS & 0.7928 & 0.3721 & 1.0000 \\
        & & OptFS   & $\textbf{0.7946}^*$ & $\textbf{0.3712}^*$ & \textbf{0.8686} \\
    \hline
\end{tabular}
}
\begin{tablenotes}
\footnotesize
\item[1] Here $*$ denotes statistically significant improvement (measured by a two-sided t-test with p-value $<0.05$) over the best baseline. \textbf{Bold} font indicates the best-performed method.
\vspace{-10pt}
\end{tablenotes}
\end{table}

In this subsection, we aim to study the influence of the OptFS method on feature interaction selection. The overall performance of our OptFS and AutoFIS on DeepFM and FM backbone models are reported in Table \ref{Table:FIS}. We summarize our observation below.

Firstly, compared with backbone models that do not perform any feature interaction selection, AutoFIS and OptFS achieve higher performance. Such an observation points out the existence of useless feature interaction on both datasets. 

Secondly, the performance of OptFS and AutoFIS differs on different models. With fewer features in the feature set, OptFS achieves nearly the same performance as AutoFIS on FM while performing significantly better on DeepFM. This is because OptFS focuses on feature-level interactions, which are more fine-grained than the field-level interactions adopted by AutoFIS.

Finally, it is also worth mentioning that OptFS can reduce $13$\% to $96$\% of features while AutoFIS is conducted on all possible features without any reduction.

\subsection{Transferability Study(RQ2)}
\label{sec:experiment_transfer}

In this subsection, we investigate the transferability of OptFS's result. The experimental settings are listed as follows. First, we search the gating vector $\mathbf{g}$ from one model, which we named the source. Then, we re-train another backbone model given the obtained gating vector, which we call the target. We study the transferability between DeepFM, DCN and IPNN backbone models over both Criteo and Avazu datasets. Based on the results shown in Table \ref{Table:TS}, we can easily observe that all transformation leads to performance degradation. Such degradation is even considered significant over the Avazu dataset. Therefore, feature interaction operations require different feature sets to achieve high performance. We can conclude that the selection of the feature set needs to incorporate the interaction operation, which further highlights the importance of selecting both features and their interactions in a unified, end-to-end trainable way.

\begin{table}[!htbp]
\centering
\caption{Transferability Analysis on Criteo and Avazu.}	\label{Table:TS}
\vspace{-5pt}
\resizebox{.40\textwidth}{!}{
\begin{tabular}{c|c|c|ccc}
    \hline
        & \multirow{2}{*}{Target} & \multirow{2}{*}{Source} & \multicolumn{3}{c}{Metrics} \\
    \cline{4-6}
        & & & AUC$\uparrow$ & Logloss$\downarrow$ & Ratio$\downarrow$ \\
    \hline
        \multirow{9}{*}{\rotatebox{90}{Criteo}} 
        & \multirow{3}{*}{DeepFM} 
        & DeepFM    & \textbf{0.8100} & \textbf{0.4415} & \textbf{0.0422} \\
        & & DCN     & 0.8097 & 0.4419 & 0.0802 \\
        & & IPNN    & 0.8097 & 0.4418 & 0.0719 \\
    \cline{2-6}
        & \multirow{3}{*}{DCN} 
        & DCN       & \textbf{0.8111} & \textbf{0.4405} & 0.0802 \\
        & & DeepFM  & 0.8106 & 0.4410 & \textbf{0.0422} \\
        & & IPNN    & 0.8107 & 0.4410 & 0.0719 \\
    \cline{2-6}
        & \multirow{3}{*}{IPNN} 
        & IPNN      & \textbf{0.8116} & \textbf{0.4401} & 0.0719 \\
        & & DCN     & 0.8113 & 0.4404 & 0.0802 \\
        & & DeepFM  & 0.8114 & 0.4403 & \textbf{0.0422} \\
    \hline
        \multirow{9}{*}{\rotatebox{90}{Avazu}} 
        & \multirow{3}{*}{DeepFM} 
        & DeepFM    & $\textbf{0.7946}^*$ & $\textbf{0.3712}^*$ & 0.8686 \\
        & & DCN     & 0.7873 & 0.3754 & \textbf{0.8665} \\
        & & IPNN    & 0.7872 & 0.3755 & 0.9118 \\
    \cline{2-6}
        & \multirow{3}{*}{DCN} 
        & DCN       & $\textbf{0.7932}^*$ & $\textbf{0.3718}^*$ & \textbf{0.8665} \\
        & & DeepFM  & 0.7879 & 0.3784 & 0.8686 \\
        & & IPNN    & 0.7860 & 0.3762 & 0.9118 \\
    \cline{2-6}
        & \multirow{3}{*}{IPNN} 
        & IPNN      & $\textbf{0.7950}^*$ & $\textbf{0.3709}^*$ & 0.9118 \\
        & & DCN     & 0.7907 & 0.3747 & \textbf{0.8665} \\
        & & DeepFM  & 0.7908 & 0.3748 & 0.8686 \\
    \hline
\end{tabular}
}
\begin{tablenotes}
\footnotesize
\item[1] Here $*$ denotes statistically significant improvement (measured by a two-sided t-test with p-value $<0.05$) over the best baseline. \textbf{Bold} font indicates the best-performed method.
\vspace{-10pt}
\end{tablenotes}
\end{table}

\subsection{Ablation Study(RQ3)}
\label{sec:experiment_ablation}

In this subsection, we conduct the ablation study over the influence of the re-training stage, which is detailedly illustrated in Section \ref{sec:method_lbc_retrain}. In Section \ref{sec:method_lbc_retrain}, we propose a \textbf{c}ustomized \textbf{i}nitialization method, namely \textit{c.i.}, during the re-training stage. Here we compare it with the other three methods of obtaining model parameters: (i) \textit{w.o.}, which is the abbreviation for \textbf{w}ith\textbf{o}ut re-training, directly inherit the model parameters from the searching stage; (ii) \textit{r.i.} \textbf{r}andomly \textbf{i}nitialize the model parameters; (iii) \textit{l.t.h.}, which stands for \textbf{l}ottery \textbf{t}icket \textbf{h}ypothesis, is a common method for re-training sparse network~\cite{LTH}. Specifically, it initializes the model parameters with the same seed from the searching stage. The experiment is conducted over three backbone models, DeepFM, DCN and IPNN, over Criteo and Avazu benchmarks. We can make the following observations based on the result shown in Table \ref{Table:AS}. 

\begin{table}[!htbp]
\centering
\caption{Ablation Study Regarding the Re-training Stage.}	\label{Table:AS}
\vspace{-5pt}
\resizebox{0.45\textwidth}{!}{
\begin{tabular}{c|c|c|cccc}
    \hline
        & \multirow{2}{*}{Model} & \multirow{2}{*}{Metrics} & \multicolumn{4}{c}{Methods} \\
    \cline{4-7}
        & & & w.o. & r.i. & l.t.h. & c.i. \\
    \hline
        \multirow{6}{*}{\rotatebox{90}{Criteo}} 
        & \multirow{2}{*}{DeepFM} 
        & AUC$\uparrow$         & 0.8012 & \textbf{0.8100} & \textbf{0.8100} & \textbf{0.8100} \\
        & & Logloss$\downarrow$ & 0.4686 & 0.4416 & \textbf{0.4415} & \textbf{0.4415} \\
    \cline{2-7}
        & \multirow{2}{*}{DCN} 
        & AUC$\uparrow$         & 0.8077 & 0.8109 & 0.8108 & \textbf{0.8111} \\
        & & Logloss$\downarrow$ & 0.4522 & 0.4407 & 0.4408 & \textbf{0.4405} \\
    \cline{2-7}
        & \multirow{2}{*}{IPNN} 
        & AUC$\uparrow$         & 0.7757 & 0.8113 & 0.8114 & \textbf{0.8116} \\
        & & Logloss$\downarrow$ & 0.4998 & 0.4404 & 0.4403 & \textbf{0.4401} \\
    \hline
        \multirow{6}{*}{\rotatebox{90}{Avazu}} &
        \multirow{2}{*}{DeepFM} 
        & AUC$\uparrow$         & 0.6972 & 0.7873 & 0.7883 & $\textbf{0.7946}^*$ \\
        & & Logloss$\downarrow$ & 0.5017 & 0.3754 & 0.3790 & $\textbf{0.3712}^*$ \\
    \cline{2-7}
        & \multirow{2}{*}{DCN} 
        & AUC$\uparrow$         & 0.7122 & 0.7870 & 0.7858 & $\textbf{0.7932}^*$ \\
        & & Logloss$\downarrow$ & 0.4736 & 0.3801 & 0.3764 & $\textbf{0.3718}^*$ \\
    \cline{2-7}
        & \multirow{2}{*}{IPNN} 
        & AUC$\uparrow$         & 0.7560 & 0.7912 & 0.7910 & $\textbf{0.7950}^*$ \\
        & & Logloss$\downarrow$ & 0.4411 & 0.3745 & 0.3745 & $\textbf{0.3709}^*$ \\
    \hline
\end{tabular}
}
\begin{tablenotes}
\footnotesize
\item[1] Here $*$ denotes statistically significant improvement (measured by a two-sided t-test with p-value $<0.05$) over the best baseline. \textbf{Bold} font indicates the best-performed method. Here \textit{w.o.} stands for \textbf{w}ith\textbf{o}ut re-training, \textit{r.i.} stands for re-training with \textbf{r}andom \textbf{i}nitialization, \textit{l.t.h.} stands for initialization using \textbf{l}ottery \textbf{t}icket \textbf{h}ypothesis~\cite{LTH}, \textit{c.i.} stands for re-training with \textbf{c}ustomized \textbf{i}nitialization, as previously discussed in Section \ref{sec:method_lbc}.
\vspace{-10pt}
\end{tablenotes}
\end{table}

Firstly, we can easily observe that re-training can improve performance regardless of its setting. Without re-training, the neural network will inherit the sub-optimal model parameters from the searching stage, which is influenced by the non-binary element in the gating vector. Re-training improves the model performance under the constraint of the gating vector.

Secondly, \textit{c.i.} constantly outperforms the other two re-training methods. Such performance gaps are considered significant on all three backbone models over the Avazu dataset. This is likely because, on the Avazu dataset, the backbone models are usually trained for only one epoch before they get early-stopped for over-fitting. Hence, it further increases the importance of initialization during the re-training stage. This observation validates the necessity of introducing customized initialization in CTR prediction.

\subsection{Efficiency Analysis(RQ4)}

In addition to model performance, efficiency is vital when deploying the CTR prediction model in reality. In this section, we investigate the time and space complexity of OptFS.

\begin{figure}[!htbp]
    \centering
    \includegraphics[width=0.35\textwidth]{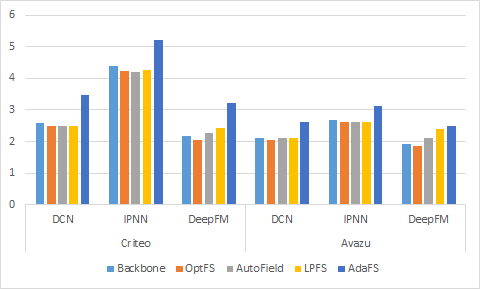}
    \vspace{-5pt}
    \caption{Inference Time on Criteo and Avazu Dataset. The Y-axis represents the influence time, measured by ms}
    \vspace{-15pt}
    \label{fig:Time}
\end{figure}

\begin{figure}[!htbp]
    \centering
    \subfigure[DeepFM]{
    \begin{minipage}[t]{0.2\textwidth}
    \centering
    \includegraphics[width=\textwidth]{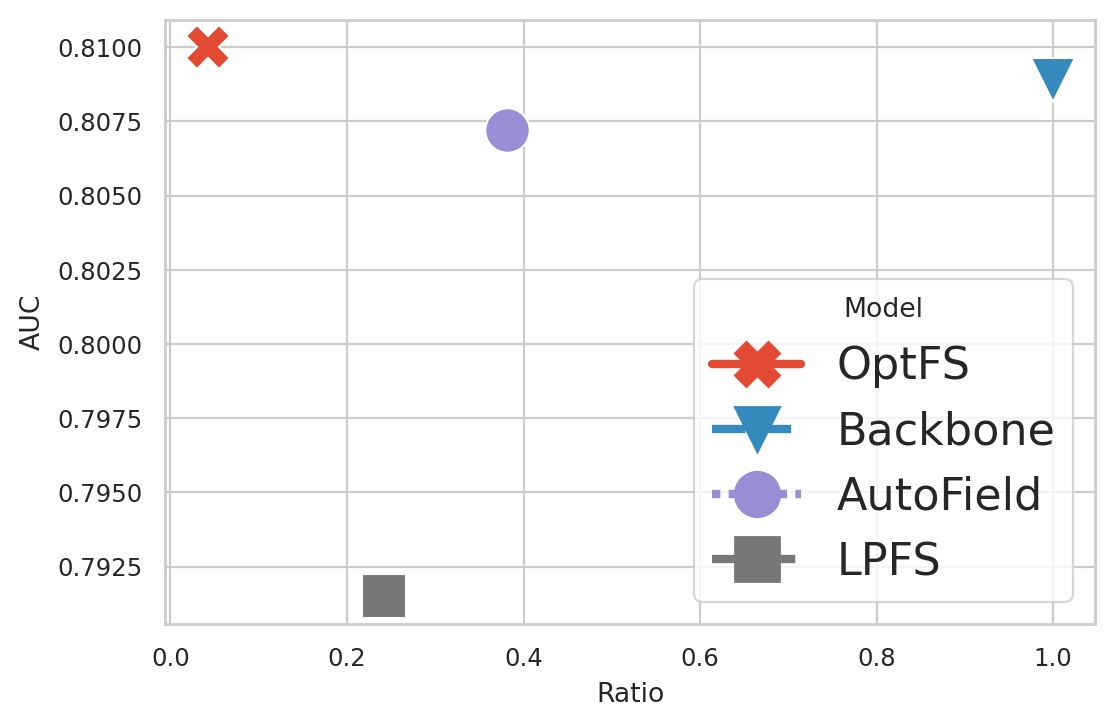}
    \vspace{-5pt}
    \label{fig:criteo-deepfm}
    \vspace{-5pt}
    \end{minipage}
    }
    \subfigure[DCN]{
    \begin{minipage}[t]{0.2\textwidth}
    \centering
    \includegraphics[width=\textwidth]{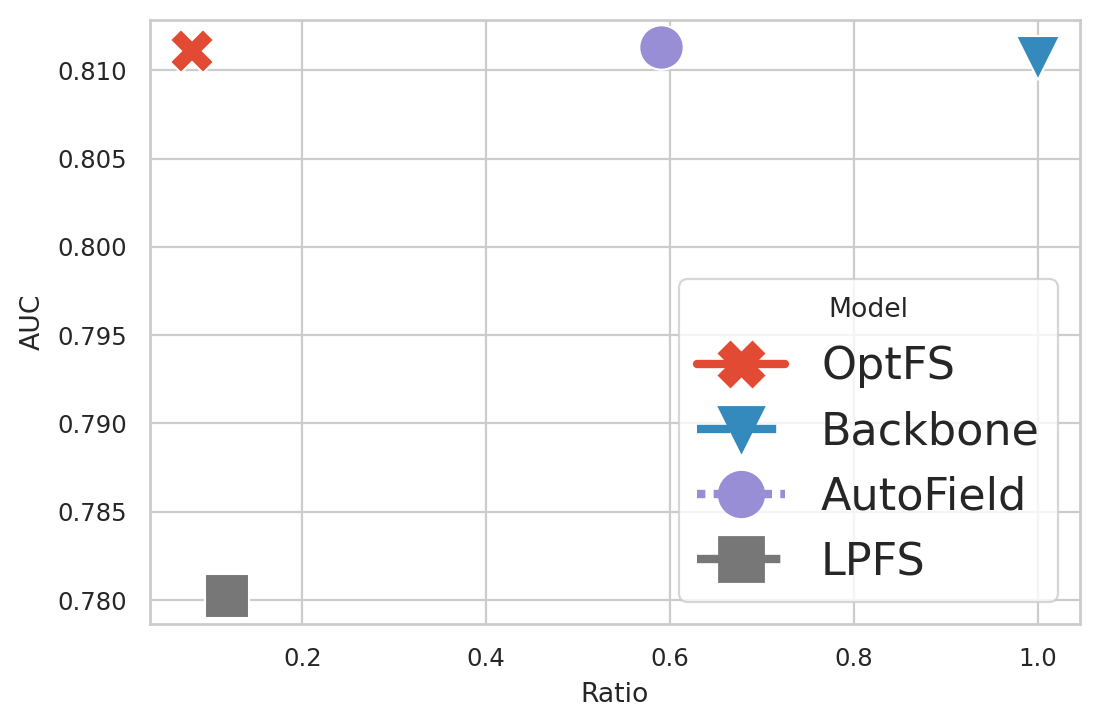}
    \vspace{-5pt}
    \label{fig:criteo-dcn}
    \vspace{-5pt}
    \end{minipage}
    }
    \vspace{-5pt}
    \caption{Visualization of efficiency-effectiveness trade-off on Criteo datasets. The closer to the top-left the better.}
    \vspace{-10pt}
    \label{fig:Param-AUC}
\end{figure}

\begin{figure*}[!htbp]
    \centering
    \subfigure[Mutual Info]{
    \begin{minipage}[t]{0.18\textwidth}
    \centering
    \includegraphics[width=\textwidth]{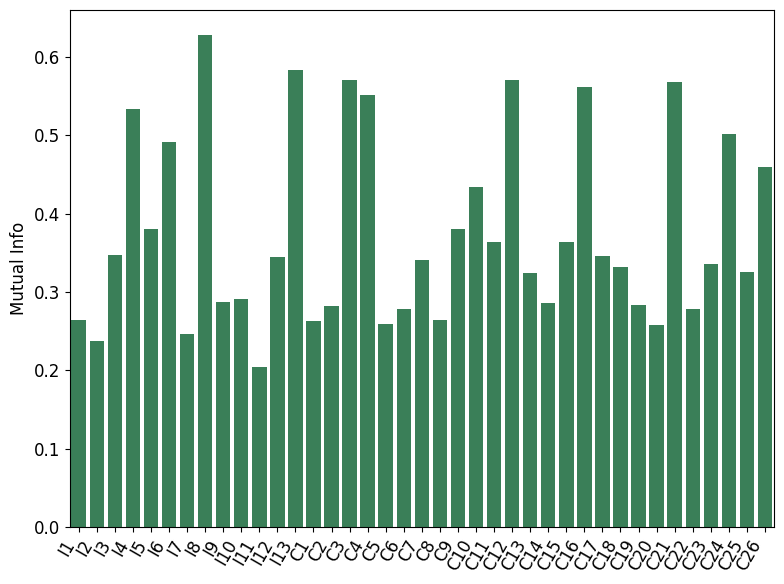}
    \vspace{-5pt}
    \label{fig:case-criteo-mi}
    \vspace{-5pt}
    \end{minipage}
    }
    \subfigure[DeepFM]{
    \begin{minipage}[t]{0.18\textwidth}
    \centering
    \includegraphics[width=\textwidth]{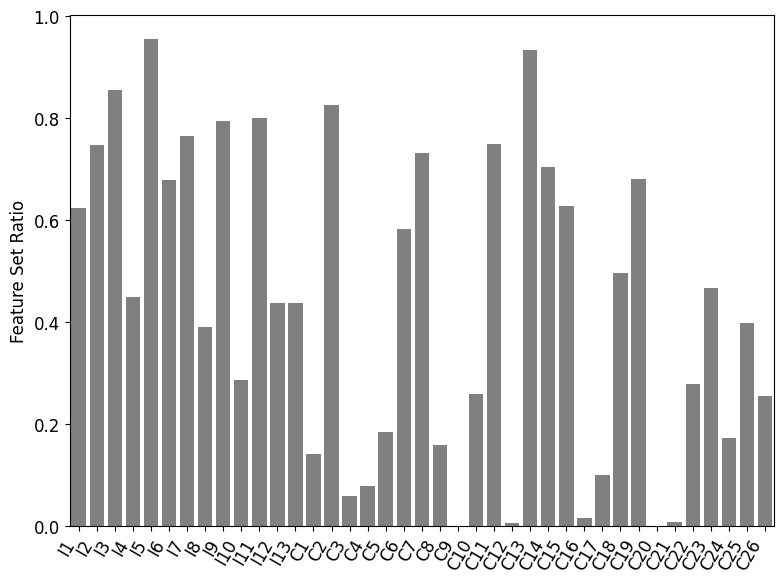}
    \vspace{-5pt}
    \label{fig:case-criteo-deepfm}
    \vspace{-5pt}
    \end{minipage}
    }
    \subfigure[DCN]{
    \begin{minipage}[t]{0.18\textwidth}
    \centering
    \includegraphics[width=\textwidth]{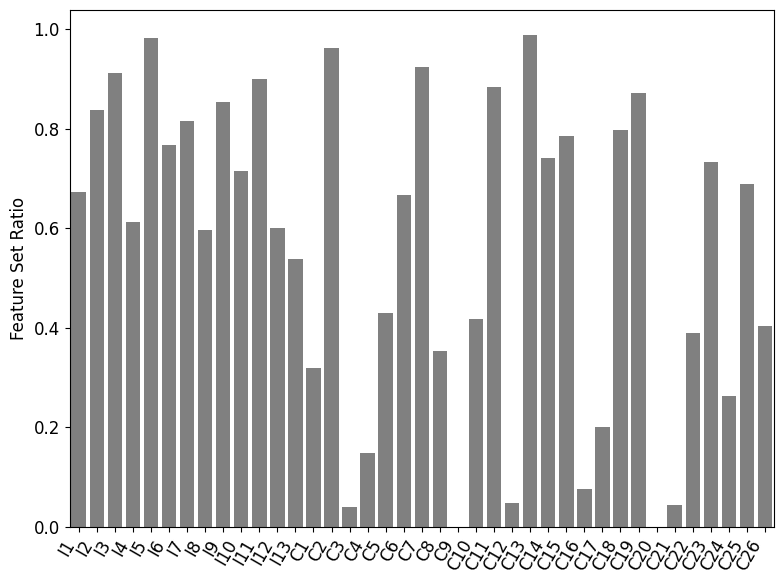}
    \vspace{-5pt}
    \label{fig:case-criteo-dcn}
    \vspace{-5pt}
    \end{minipage}
    }
    \subfigure[IPNN]{
    \begin{minipage}[t]{0.18\textwidth}
    \centering
    \includegraphics[width=\textwidth]{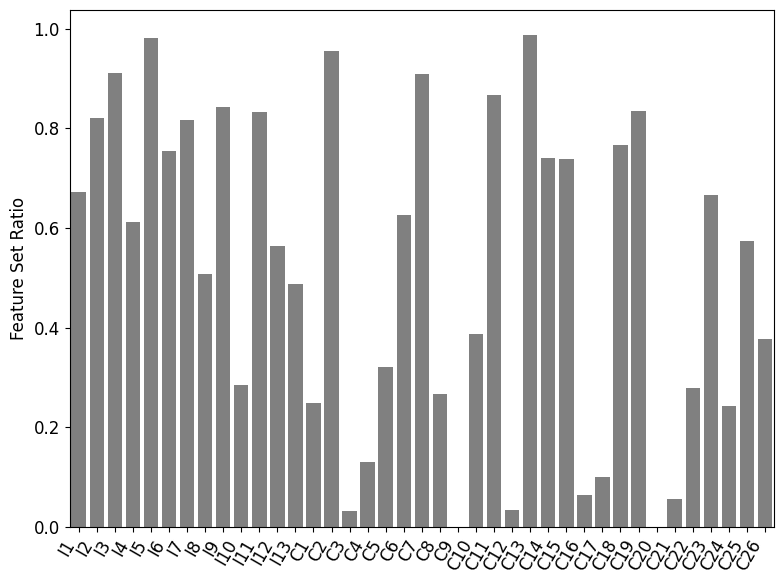}
    \vspace{-5pt}
    \label{fig:case-criteo-ipnn}
    \vspace{-5pt}
    \end{minipage}
    }
    \vspace{-10pt}
    \caption{A Case Study of OptFS output on Criteo. In all subfigures, the X-axis indicates the field identifiers. Subfigure (a) plots the mutual information scores, while subfigures (b), (c) and (d) plot the feature set ratio of OptFS on DeepFM, DCN and IPNN.}
    \vspace{-10pt}
    \label{fig:case}
\end{figure*}

\subsubsection{Time Complexity}
The inference time is crucial when deploying the model into online web systems. We define inference time as the time for inferencing one batch. The result is obtained by averaging the inference time over all batches on the validation set.

As shown in Figure \ref{fig:Time}, OptFS achieves the least inference time. This is because the feature set obtained by OptFS usually has the least features. Meanwhile, AdaFS requires the longest inference time, even longer than the backbone model. This is because it needs to determine whether keep or drop each feature dynamically during run-time.

\subsubsection{Space Complexity}
We plot the Feature Ratio-AUC curve of the DeepFM, DCN and IPNN model on the Criteo datasets in Figure \ref{fig:Param-AUC}, which reflects the relationship between the space complexity of the feature set and model performance. Notes that LPFS, AutoField and OptFS are methods that primarily aim to improve model performance. These methods have no guarantee over the final feature ratios. Hence we only plot one point for each method in the figure.

From Figure \ref{fig:Param-AUC} we can make the following observations: (i) OptFS outperforms all other baselines with the highest AUC score and the least number of features. (ii) The model performance of AutoField is comparable with OptFS and Backbone. However, given it only selects the feature set on field-level, its feature ratio tends to be higher than OptFS. (iii) The performance of LPFS is much lower than other methods.
\vspace{-5pt}

\subsection{Case Study(RQ5)}

This subsection uses a case study to investigate the optimal feature set obtained from OptFS. In Figure \ref{fig:case}, we plot the mutual information with the feature ratio on each field. For field $\mathbf{z}_i = \{\mathbf{x}_{k_i}\}$ and ground truth labels $\mathbf{y}$ ($y \in \mathbf{y}$), the mutual information between them is defined as:
\begin{equation}
    \mathbf{MI}(\mathbf{x}_{k_i},\mathbf{y}) = -\sum \mathbf{P}(y) \log \mathbf{P}(y) + \sum \mathbf{P}(\mathbf{x}_{k_i} ,y) \log \mathbf{P}(y|\mathbf{x}_{k_i}),
\end{equation}
where the first term is the marginal entropy and the second term is the conditional entropy of ground truth labels $\mathbf{y}$ given field $\mathbf{z}_i = \{\mathbf{x}_{k_i}\}$. Note that fields with high mutual information scores are more informative (hence more important) to the prediction. 

As a case study, we investigate the feature ratio for each field, shown in Figure \ref{fig:case}. We select the result from DeepFM, DCN and IPNN on the Criteo dataset. Figure \ref{fig:case-criteo-mi} shows the mutual information scores of each field, which represents how informative each field is in predicting the label. Figure \ref{fig:case-criteo-deepfm}, \ref{fig:case-criteo-dcn} and \ref{fig:case-criteo-ipnn} shows the feature ratio given each fields. As can be seen, fields with higher mutual information scores are likely to keep more features in the feature set, which indicates that OptFS obtains the optimal feature set from the field perspective.

\vspace{-3pt}

%% file: section/related_work.tex
\section{related work}
\label{sec:rw}

In this section, we review the related work. Optimizing feature set is related two topics, feature selection and feature interaction selection. The training scheme of proposed OptFS is related to learning by continuation. Thus we summarize the related work in following two subsection.

\vspace{-3pt}

\subsection{Feature and Feature Interaction Selection}
Feature selection is a key component for prediction task~\cite{DFS}. Several methods have been proposed~\cite{LASSO, MCF, AutoField, LPFS, AdaFS} to conduct feature selection for CTR models. Traditional methods~\cite{LASSO, MCF} exploit the statistical metrics of different feature fields and conduct feature field selection. Inspired by neural architecture search (NAS)~\cite{DARTS, NAO} and smoothed-$\mathcal{l}_0$ optimization respectively, AutoField~\cite{AutoField} and LPFS~\cite{LPFS} determine the selection of feature fields automatically. AdaFS~\cite{AdaFS} proposes a novel controller network to decide feature fields for each sample, which fits the dynamic recommendation.
Feature interaction selection is often employed to enhance the prediction. Some methods~\cite{AutoFIS, AutoFeature} model the problem as NAS to exploit the field-level interaction space. OptInter~\cite{OptInter} investigates the way to do feature interaction. AutoCross~\cite{AutoCross} targets on tabular data and iterative finds feature interaction based on locally optimized feature set.
We first highlight the feature set optimization problem in CTR prediction, and OptFS is different from previous methods by solving both problems in a unified manner.

\vspace{-5pt}

\subsection{Learning by Continuation}
Continuation methods are commonly used to approximate intractable optimization problems by gradually increasing the difficulty of the underlying objective. By adopting gradual relaxations to binary problems, gumbel-softmax~\cite{Gumbel-Softmax} is used to back-propagate errors during the architecture search ~\cite{fbnet} and spatial feature sparsification~\cite{feature_spar}. Other methods~\cite{Cont_Spar, DST, Grow_Spar} introduce continuous sparsification framework to speed up neural network pruning and ticket search.
OptFS adopts the learning-by-continuation scheme to effectively explore the huge feature-level search space.

\vspace{-5pt}

%% file: section/conclusion.tex
\section{conclusion}
\label{sec:conclusion}
This paper first distinguishes the feature set optimization problem. Such a problem unifies two mutually influencing questions: the selection of features and feature interactions. To our knowledge, no previous work considers these two questions uniformly. Besides, we also upgrade the granularity of the problem from field-level to feature-level. To solve such the feature set optimization problem efficiently, we propose a novel method named OptFS, which assigns a gating value to each feature for its usefulness and adopt a learning-by-continuation approach for efficient optimization. Extensive experiments on three large-scale datasets demonstrate the superiority of OptFS in model performance and feature reduction. Several ablation studies also illustrate the necessity of our design. Moreover, we also interpret the obtained result on feature fields and their interactions, highlighting that our method properly solves the feature set optimization problem.

%% file: main.bbl

\begin{thebibliography}{38}


\ifx \showCODEN    \undefined \def \showCODEN     #1{\unskip}     \fi
\ifx \showDOI      \undefined \def \showDOI       #1{#1}\fi
\ifx \showISBNx    \undefined \def \showISBNx     #1{\unskip}     \fi
\ifx \showISBNxiii \undefined \def \showISBNxiii  #1{\unskip}     \fi
\ifx \showISSN     \undefined \def \showISSN      #1{\unskip}     \fi
\ifx \showLCCN     \undefined \def \showLCCN      #1{\unskip}     \fi
\ifx \shownote     \undefined \def \shownote      #1{#1}          \fi
\ifx \showarticletitle \undefined \def \showarticletitle #1{#1}   \fi
\ifx \showURL      \undefined \def \showURL       {\relax}        \fi
\providecommand\bibfield[2]{#2}
\providecommand\bibinfo[2]{#2}
\providecommand\natexlab[1]{#1}
\providecommand\showeprint[2][]{arXiv:#2}

\bibitem[\protect\citeauthoryear{Bengio, Courville, and Vincent}{Bengio
  et~al\mbox{.}}{2013}]%
        {RLReview}
\bibfield{author}{\bibinfo{person}{Yoshua Bengio}, \bibinfo{person}{Aaron~C.
  Courville}, {and} \bibinfo{person}{Pascal Vincent}.}
  \bibinfo{year}{2013}\natexlab{}.
\newblock \showarticletitle{Representation Learning: {A} Review and New
  Perspectives}.
\newblock \bibinfo{journal}{\emph{{IEEE} Trans. Pattern Anal. Mach. Intell.}}
  \bibinfo{volume}{35}, \bibinfo{number}{8} (\bibinfo{year}{2013}),
  \bibinfo{pages}{1798--1828}.
\newblock
\urldef\tempurl%
\url{https://doi.org/10.1109/TPAMI.2013.50}
\showDOI{\tempurl}


\bibitem[\protect\citeauthoryear{Chapelle, Manavoglu, and Rosales}{Chapelle
  et~al\mbox{.}}{2015}]%
        {ADS}
\bibfield{author}{\bibinfo{person}{Olivier Chapelle}, \bibinfo{person}{Eren
  Manavoglu}, {and} \bibinfo{person}{Romer Rosales}.}
  \bibinfo{year}{2015}\natexlab{}.
\newblock \showarticletitle{Simple and Scalable Response Prediction for Display
  Advertising}.
\newblock \bibinfo{journal}{\emph{ACM Trans. Intell. Syst. Technol.}}
  \bibinfo{volume}{5}, \bibinfo{number}{4} (\bibinfo{date}{dec}
  \bibinfo{year}{2015}), \bibinfo{pages}{61}.
\newblock
\showISSN{2157-6904}


\bibitem[\protect\citeauthoryear{Don{\`{a}} and Gallinari}{Don{\`{a}} and
  Gallinari}{2021}]%
        {DFS}
\bibfield{author}{\bibinfo{person}{J{\'{e}}r{\'{e}}mie Don{\`{a}}} {and}
  \bibinfo{person}{Patrick Gallinari}.} \bibinfo{year}{2021}\natexlab{}.
\newblock \showarticletitle{Differentiable Feature Selection, {A}
  Reparameterization Approach}. In \bibinfo{booktitle}{\emph{Machine Learning
  and Knowledge Discovery in Databases. Research Track - European Conference,
  {ECML} {PKDD} 2021}} \emph{(\bibinfo{series}{Lecture Notes in Computer
  Science}, Vol.~\bibinfo{volume}{12977})}. \bibinfo{publisher}{Springer},
  \bibinfo{address}{Bilbao, Spain}, \bibinfo{pages}{414--429}.
\newblock
\urldef\tempurl%
\url{https://doi.org/10.1007/978-3-030-86523-8\_25}
\showDOI{\tempurl}


\bibitem[\protect\citeauthoryear{Frankle and Carbin}{Frankle and
  Carbin}{2019}]%
        {LTH}
\bibfield{author}{\bibinfo{person}{Jonathan Frankle} {and}
  \bibinfo{person}{Michael Carbin}.} \bibinfo{year}{2019}\natexlab{}.
\newblock \showarticletitle{The Lottery Ticket Hypothesis: Finding Sparse,
  Trainable Neural Networks}. In \bibinfo{booktitle}{\emph{7th International
  Conference on Learning Representations, {ICLR} 2019}}.
  \bibinfo{publisher}{OpenReview.net}, \bibinfo{address}{New Orleans, LA, USA}.
\newblock


\bibitem[\protect\citeauthoryear{Glorot and Bengio}{Glorot and Bengio}{2010}]%
        {Xavier}
\bibfield{author}{\bibinfo{person}{Xavier Glorot} {and} \bibinfo{person}{Yoshua
  Bengio}.} \bibinfo{year}{2010}\natexlab{}.
\newblock \showarticletitle{Understanding the difficulty of training deep
  feedforward neural networks}. In \bibinfo{booktitle}{\emph{13th International
  Conference on Artificial Intelligence and Statistics, {AISTATS} 2010}}
  \emph{(\bibinfo{series}{{JMLR} Proceedings}, Vol.~\bibinfo{volume}{9})}.
  \bibinfo{publisher}{JMLR.org}, \bibinfo{address}{Italy},
  \bibinfo{pages}{249--256}.
\newblock


\bibitem[\protect\citeauthoryear{Guo, Guo, Gao, Tang, He, and Liu}{Guo
  et~al\mbox{.}}{2021}]%
        {sfctr}
\bibfield{author}{\bibinfo{person}{Huifeng Guo}, \bibinfo{person}{Wei Guo},
  \bibinfo{person}{Yong Gao}, \bibinfo{person}{Ruiming Tang},
  \bibinfo{person}{Xiuqiang He}, {and} \bibinfo{person}{Wenzhi Liu}.}
  \bibinfo{year}{2021}\natexlab{}.
\newblock \showarticletitle{ScaleFreeCTR: MixCache-based Distributed Training
  System for {CTR} Models with Huge Embedding Table}. In
  \bibinfo{booktitle}{\emph{{SIGIR} '21: The 44th International {ACM} {SIGIR}
  Conference on Research and Development in Information Retrieval}}.
  \bibinfo{publisher}{{ACM}}, \bibinfo{address}{Virtual Event, Canada},
  \bibinfo{pages}{1269--1278}.
\newblock
\urldef\tempurl%
\url{https://doi.org/10.1145/3404835.3462976}
\showDOI{\tempurl}


\bibitem[\protect\citeauthoryear{Guo, Tang, Ye, Li, and He}{Guo
  et~al\mbox{.}}{2017}]%
        {DeepFM}
\bibfield{author}{\bibinfo{person}{Huifeng Guo}, \bibinfo{person}{Ruiming
  Tang}, \bibinfo{person}{Yunming Ye}, \bibinfo{person}{Zhenguo Li}, {and}
  \bibinfo{person}{Xiuqiang He}.} \bibinfo{year}{2017}\natexlab{}.
\newblock \showarticletitle{DeepFM: {A} Factorization-Machine based Neural
  Network for {CTR} Prediction}. In \bibinfo{booktitle}{\emph{26th
  International Joint Conference on Artificial Intelligence, {IJCAI} 2017}}.
  \bibinfo{publisher}{ijcai.org}, \bibinfo{address}{Melbourne, Australia},
  \bibinfo{pages}{1725--1731}.
\newblock


\bibitem[\protect\citeauthoryear{Guo, Liu, Tan, Liao, Chang, Liu, Yang, Liu,
  Kong, Chen, and Song}{Guo et~al\mbox{.}}{2022}]%
        {LPFS}
\bibfield{author}{\bibinfo{person}{Yi Guo}, \bibinfo{person}{Zhaocheng Liu},
  \bibinfo{person}{Jianchao Tan}, \bibinfo{person}{Chao Liao},
  \bibinfo{person}{Daqing Chang}, \bibinfo{person}{Qiang Liu},
  \bibinfo{person}{Sen Yang}, \bibinfo{person}{Ji Liu},
  \bibinfo{person}{Dongying Kong}, \bibinfo{person}{Zhi Chen}, {and}
  \bibinfo{person}{Chengru Song}.} \bibinfo{year}{2022}\natexlab{}.
\newblock \showarticletitle{{LPFS:} Learnable Polarizing Feature Selection for
  Click-Through Rate Prediction}.
\newblock \bibinfo{journal}{\emph{CoRR}}  \bibinfo{volume}{abs/2206.00267}
  (\bibinfo{year}{2022}).
\newblock
\urldef\tempurl%
\url{https://doi.org/10.48550/arXiv.2206.00267}
\showDOI{\tempurl}
\showeprint[arXiv]{2206.00267}


\bibitem[\protect\citeauthoryear{Hastie, Tibshirani, and Friedman}{Hastie
  et~al\mbox{.}}{2009}]%
        {Elements_SL}
\bibfield{author}{\bibinfo{person}{Trevor Hastie}, \bibinfo{person}{Robert
  Tibshirani}, {and} \bibinfo{person}{Jerome~H. Friedman}.}
  \bibinfo{year}{2009}\natexlab{}.
\newblock \bibinfo{booktitle}{\emph{The Elements of Statistical Learning: Data
  Mining, Inference, and Prediction, 2nd Edition}}.
\newblock \bibinfo{publisher}{Springer}, \bibinfo{address}{Berlin, Germany}.
\newblock
\showISBNx{9780387848570}
\urldef\tempurl%
\url{https://doi.org/10.1007/978-0-387-84858-7}
\showDOI{\tempurl}


\bibitem[\protect\citeauthoryear{Ioffe and Szegedy}{Ioffe and Szegedy}{2015}]%
        {BatchNorm}
\bibfield{author}{\bibinfo{person}{Sergey Ioffe} {and}
  \bibinfo{person}{Christian Szegedy}.} \bibinfo{year}{2015}\natexlab{}.
\newblock \showarticletitle{Batch Normalization: Accelerating Deep Network
  Training by Reducing Internal Covariate Shift}. In
  \bibinfo{booktitle}{\emph{32nd International Conference on Machine Learning,
  {ICML} 2015}} \emph{(\bibinfo{series}{{JMLR} Workshop and Conference
  Proceedings}, Vol.~\bibinfo{volume}{37})}. \bibinfo{publisher}{JMLR.org},
  \bibinfo{address}{France}, \bibinfo{pages}{448--456}.
\newblock


\bibitem[\protect\citeauthoryear{Jang, Gu, and Poole}{Jang
  et~al\mbox{.}}{2017}]%
        {Gumbel-Softmax}
\bibfield{author}{\bibinfo{person}{Eric Jang}, \bibinfo{person}{Shixiang Gu},
  {and} \bibinfo{person}{Ben Poole}.} \bibinfo{year}{2017}\natexlab{}.
\newblock \showarticletitle{Categorical Reparameterization with
  Gumbel-Softmax}. In \bibinfo{booktitle}{\emph{5th International Conference on
  Learning Representations, {ICLR} 2017}}. \bibinfo{publisher}{OpenReview.net},
  \bibinfo{address}{Toulon, France}.
\newblock


\bibitem[\protect\citeauthoryear{Khawar, Hang, Tang, Liu, Li, and He}{Khawar
  et~al\mbox{.}}{2020}]%
        {AutoFeature}
\bibfield{author}{\bibinfo{person}{Farhan Khawar}, \bibinfo{person}{Xu Hang},
  \bibinfo{person}{Ruiming Tang}, \bibinfo{person}{Bin Liu},
  \bibinfo{person}{Zhenguo Li}, {and} \bibinfo{person}{Xiuqiang He}.}
  \bibinfo{year}{2020}\natexlab{}.
\newblock \showarticletitle{AutoFeature: Searching for Feature Interactions and
  Their Architectures for Click-through Rate Prediction}. In
  \bibinfo{booktitle}{\emph{{CIKM} '20: The 29th {ACM} International Conference
  on Information and Knowledge Management}}. \bibinfo{publisher}{{ACM}},
  \bibinfo{address}{Virtual Event, Ireland}, \bibinfo{pages}{625--634}.
\newblock
\urldef\tempurl%
\url{https://doi.org/10.1145/3340531.3411912}
\showDOI{\tempurl}


\bibitem[\protect\citeauthoryear{Lin, Zhao, Wang, Xu, and Wu}{Lin
  et~al\mbox{.}}{2022}]%
        {AdaFS}
\bibfield{author}{\bibinfo{person}{Weilin Lin}, \bibinfo{person}{Xiangyu Zhao},
  \bibinfo{person}{Yejing Wang}, \bibinfo{person}{Tong Xu}, {and}
  \bibinfo{person}{Xian Wu}.} \bibinfo{year}{2022}\natexlab{}.
\newblock \showarticletitle{AdaFS: Adaptive Feature Selection in Deep
  Recommender System}. In \bibinfo{booktitle}{\emph{{KDD} '22: The 28th {ACM}
  {SIGKDD} Conference on Knowledge Discovery and Data Mining}}.
  \bibinfo{publisher}{{ACM}}, \bibinfo{address}{Washington, DC, USA},
  \bibinfo{pages}{3309--3317}.
\newblock
\urldef\tempurl%
\url{https://doi.org/10.1145/3534678.3539204}
\showDOI{\tempurl}


\bibitem[\protect\citeauthoryear{Liu, Zhu, Li, Zhang, Lai, Tang, He, Li, and
  Yu}{Liu et~al\mbox{.}}{2020b}]%
        {AutoFIS}
\bibfield{author}{\bibinfo{person}{Bin Liu}, \bibinfo{person}{Chenxu Zhu},
  \bibinfo{person}{Guilin Li}, \bibinfo{person}{Weinan Zhang},
  \bibinfo{person}{Jincai Lai}, \bibinfo{person}{Ruiming Tang},
  \bibinfo{person}{Xiuqiang He}, \bibinfo{person}{Zhenguo Li}, {and}
  \bibinfo{person}{Yong Yu}.} \bibinfo{year}{2020}\natexlab{b}.
\newblock \showarticletitle{AutoFIS: Automatic Feature Interaction Selection in
  Factorization Models for Click-Through Rate Prediction}. In
  \bibinfo{booktitle}{\emph{{KDD} '20: The 26th {ACM} {SIGKDD} Conference on
  Knowledge Discovery and Data Mining}}. \bibinfo{publisher}{{ACM}},
  \bibinfo{address}{Virtual Event, CA, USA}, \bibinfo{pages}{2636--2645}.
\newblock
\urldef\tempurl%
\url{https://doi.org/10.1145/3394486.3403314}
\showDOI{\tempurl}


\bibitem[\protect\citeauthoryear{Liu, Simonyan, and Yang}{Liu
  et~al\mbox{.}}{2019}]%
        {DARTS}
\bibfield{author}{\bibinfo{person}{Hanxiao Liu}, \bibinfo{person}{Karen
  Simonyan}, {and} \bibinfo{person}{Yiming Yang}.}
  \bibinfo{year}{2019}\natexlab{}.
\newblock \showarticletitle{{DARTS:} Differentiable Architecture Search}. In
  \bibinfo{booktitle}{\emph{7th International Conference on Learning
  Representations, {ICLR} 2019}}. \bibinfo{publisher}{OpenReview.net},
  \bibinfo{address}{USA}.
\newblock


\bibitem[\protect\citeauthoryear{Liu, Xu, Shi, Cheung, and So}{Liu
  et~al\mbox{.}}{2020a}]%
        {DST}
\bibfield{author}{\bibinfo{person}{Junjie Liu}, \bibinfo{person}{Zhe Xu},
  \bibinfo{person}{Runbin Shi}, \bibinfo{person}{Ray C.~C. Cheung}, {and}
  \bibinfo{person}{Hayden~Kwok{-}Hay So}.} \bibinfo{year}{2020}\natexlab{a}.
\newblock \showarticletitle{Dynamic Sparse Training: Find Efficient Sparse
  Network From Scratch With Trainable Masked Layers}. In
  \bibinfo{booktitle}{\emph{8th International Conference on Learning
  Representations, {ICLR} 2020}}. \bibinfo{publisher}{OpenReview.net},
  \bibinfo{address}{Addis Ababa, Ethiopia}.
\newblock


\bibitem[\protect\citeauthoryear{Liu, Liu, Zhang, Chen, and Zhu}{Liu
  et~al\mbox{.}}{2021}]%
        {MCF}
\bibfield{author}{\bibinfo{person}{Qiang Liu}, \bibinfo{person}{Zhaocheng Liu},
  \bibinfo{person}{Haoli Zhang}, \bibinfo{person}{Yuntian Chen}, {and}
  \bibinfo{person}{Jun Zhu}.} \bibinfo{year}{2021}\natexlab{}.
\newblock \showarticletitle{Mining Cross Features for Financial Credit Risk
  Assessment}. In \bibinfo{booktitle}{\emph{{CIKM} '21: The 30th {ACM}
  International Conference on Information and Knowledge Management}}.
  \bibinfo{publisher}{{ACM}}, \bibinfo{address}{Virtual Event, Queensland,
  Australia}, \bibinfo{pages}{1069--1078}.
\newblock
\urldef\tempurl%
\url{https://doi.org/10.1145/3459637.3482371}
\showDOI{\tempurl}


\bibitem[\protect\citeauthoryear{Luo, Tian, Qin, Chen, and Liu}{Luo
  et~al\mbox{.}}{2018}]%
        {NAO}
\bibfield{author}{\bibinfo{person}{Renqian Luo}, \bibinfo{person}{Fei Tian},
  \bibinfo{person}{Tao Qin}, \bibinfo{person}{Enhong Chen}, {and}
  \bibinfo{person}{Tie{-}Yan Liu}.} \bibinfo{year}{2018}\natexlab{}.
\newblock \showarticletitle{Neural Architecture Optimization}. In
  \bibinfo{booktitle}{\emph{31st Annual Conference on Neural Information
  Processing Systems 2018, NeurIPS 2018}}. \bibinfo{publisher}{Curran
  Associates}, \bibinfo{address}{Montr{\'{e}}al, Canada},
  \bibinfo{pages}{7827--7838}.
\newblock


\bibitem[\protect\citeauthoryear{Luo, Wang, Zhou, Yao, Tu, Chen, Dai, and
  Yang}{Luo et~al\mbox{.}}{2019}]%
        {AutoCross}
\bibfield{author}{\bibinfo{person}{Yuanfei Luo}, \bibinfo{person}{Mengshuo
  Wang}, \bibinfo{person}{Hao Zhou}, \bibinfo{person}{Quanming Yao},
  \bibinfo{person}{Wei{-}Wei Tu}, \bibinfo{person}{Yuqiang Chen},
  \bibinfo{person}{Wenyuan Dai}, {and} \bibinfo{person}{Qiang Yang}.}
  \bibinfo{year}{2019}\natexlab{}.
\newblock \showarticletitle{AutoCross: Automatic Feature Crossing for Tabular
  Data in Real-World Applications}. In \bibinfo{booktitle}{\emph{25th {ACM}
  International Conference on Knowledge Discovery {\&} Data Mining, {KDD}
  2019}}. \bibinfo{publisher}{{ACM}}, \bibinfo{address}{Anchorage, AK, USA},
  \bibinfo{pages}{1936--1945}.
\newblock
\urldef\tempurl%
\url{https://doi.org/10.1145/3292500.3330679}
\showDOI{\tempurl}


\bibitem[\protect\citeauthoryear{Lyu, Tang, Guo, Tang, He, Zhang, and Liu}{Lyu
  et~al\mbox{.}}{2022a}]%
        {OptInter}
\bibfield{author}{\bibinfo{person}{Fuyuan Lyu}, \bibinfo{person}{Xing Tang},
  \bibinfo{person}{Huifeng Guo}, \bibinfo{person}{Ruiming Tang},
  \bibinfo{person}{Xiuqiang He}, \bibinfo{person}{Rui Zhang}, {and}
  \bibinfo{person}{Xue Liu}.} \bibinfo{year}{2022}\natexlab{a}.
\newblock \showarticletitle{Memorize, Factorize, or be Naive: Learning Optimal
  Feature Interaction Methods for {CTR} Prediction}. In
  \bibinfo{booktitle}{\emph{38th {IEEE} International Conference on Data
  Engineering, {ICDE} 2022}}. \bibinfo{publisher}{{IEEE}},
  \bibinfo{address}{Kuala Lumpur, Malaysia}, \bibinfo{pages}{1450--1462}.
\newblock
\urldef\tempurl%
\url{https://doi.org/10.1109/ICDE53745.2022.00113}
\showDOI{\tempurl}


\bibitem[\protect\citeauthoryear{Lyu, Tang, Zhu, Guo, Zhang, Tang, and Liu}{Lyu
  et~al\mbox{.}}{2022b}]%
        {OptEmbed}
\bibfield{author}{\bibinfo{person}{Fuyuan Lyu}, \bibinfo{person}{Xing Tang},
  \bibinfo{person}{Hong Zhu}, \bibinfo{person}{Huifeng Guo},
  \bibinfo{person}{Yingxue Zhang}, \bibinfo{person}{Ruiming Tang}, {and}
  \bibinfo{person}{Xue Liu}.} \bibinfo{year}{2022}\natexlab{b}.
\newblock \showarticletitle{OptEmbed: Learning Optimal Embedding Table for
  Click-through Rate Prediction}.
\newblock \bibinfo{journal}{\emph{CoRR}}  \bibinfo{volume}{abs/2208.04482}
  (\bibinfo{year}{2022}).
\newblock


\bibitem[\protect\citeauthoryear{Meng, Zhang, Li, Li, Zhu, and Sun}{Meng
  et~al\mbox{.}}{2021}]%
        {AutoPI}
\bibfield{author}{\bibinfo{person}{Ze Meng}, \bibinfo{person}{Jinnian Zhang},
  \bibinfo{person}{Yumeng Li}, \bibinfo{person}{Jiancheng Li},
  \bibinfo{person}{Tanchao Zhu}, {and} \bibinfo{person}{Lifeng Sun}.}
  \bibinfo{year}{2021}\natexlab{}.
\newblock \showarticletitle{A General Method For Automatic Discovery of
  Powerful Interactions In Click-Through Rate Prediction}. In
  \bibinfo{booktitle}{\emph{{SIGIR} '21: The 44th International {ACM} {SIGIR}
  Conference on Research and Development in Information Retrieval}}.
  \bibinfo{publisher}{{ACM}}, \bibinfo{address}{Canada},
  \bibinfo{pages}{1298--1307}.
\newblock


\bibitem[\protect\citeauthoryear{Natarajan}{Natarajan}{1995}]%
        {l2norm}
\bibfield{author}{\bibinfo{person}{B.~K. Natarajan}.}
  \bibinfo{year}{1995}\natexlab{}.
\newblock \showarticletitle{Sparse Approximate Solutions to Linear Systems}.
\newblock \bibinfo{journal}{\emph{{SIAM} J. Comput.}} \bibinfo{volume}{24},
  \bibinfo{number}{2} (\bibinfo{year}{1995}), \bibinfo{pages}{227--234}.
\newblock
\urldef\tempurl%
\url{https://doi.org/10.1137/S0097539792240406}
\showDOI{\tempurl}


\bibitem[\protect\citeauthoryear{Qu, Cai, Ren, Zhang, Yu, Wen, and Wang}{Qu
  et~al\mbox{.}}{2016}]%
        {IPNN}
\bibfield{author}{\bibinfo{person}{Yanru Qu}, \bibinfo{person}{Han Cai},
  \bibinfo{person}{Kan Ren}, \bibinfo{person}{Weinan Zhang},
  \bibinfo{person}{Yong Yu}, \bibinfo{person}{Ying Wen}, {and}
  \bibinfo{person}{Jun Wang}.} \bibinfo{year}{2016}\natexlab{}.
\newblock \showarticletitle{Product-Based Neural Networks for User Response
  Prediction}. In \bibinfo{booktitle}{\emph{2016 IEEE 16th International
  Conference on Data Mining (ICDM)}}. \bibinfo{publisher}{IEEE},
  \bibinfo{address}{Barcelona, Spain}, \bibinfo{pages}{1149--1154}.
\newblock
\urldef\tempurl%
\url{https://doi.org/10.1109/ICDM.2016.0151}
\showDOI{\tempurl}


\bibitem[\protect\citeauthoryear{Qu, Fang, Zhang, Tang, Niu, Guo, Yu, and
  He}{Qu et~al\mbox{.}}{2019}]%
        {PIN}
\bibfield{author}{\bibinfo{person}{Yanru Qu}, \bibinfo{person}{Bohui Fang},
  \bibinfo{person}{Weinan Zhang}, \bibinfo{person}{Ruiming Tang},
  \bibinfo{person}{Minzhe Niu}, \bibinfo{person}{Huifeng Guo},
  \bibinfo{person}{Yong Yu}, {and} \bibinfo{person}{Xiuqiang He}.}
  \bibinfo{year}{2019}\natexlab{}.
\newblock \showarticletitle{Product-Based Neural Networks for User Response
  Prediction over Multi-Field Categorical Data}.
\newblock \bibinfo{journal}{\emph{{ACM} Trans. Inf. Syst.}}
  \bibinfo{volume}{37}, \bibinfo{number}{1} (\bibinfo{year}{2019}),
  \bibinfo{pages}{5:1--5:35}.
\newblock


\bibitem[\protect\citeauthoryear{Rendle}{Rendle}{2010}]%
        {FM}
\bibfield{author}{\bibinfo{person}{Steffen Rendle}.}
  \bibinfo{year}{2010}\natexlab{}.
\newblock \showarticletitle{Factorization Machines}. In
  \bibinfo{booktitle}{\emph{{ICDM} 2010, The 10th {IEEE} International
  Conference on Data Mining}}. \bibinfo{publisher}{{IEEE} Computer Society},
  \bibinfo{address}{Sydney, Australia}, \bibinfo{pages}{995--1000}.
\newblock


\bibitem[\protect\citeauthoryear{Richardson, Dominowska, and Ragno}{Richardson
  et~al\mbox{.}}{2007}]%
        {LR}
\bibfield{author}{\bibinfo{person}{Matthew Richardson}, \bibinfo{person}{Ewa
  Dominowska}, {and} \bibinfo{person}{Robert Ragno}.}
  \bibinfo{year}{2007}\natexlab{}.
\newblock \showarticletitle{Predicting clicks: estimating the click-through
  rate for new ads}. In \bibinfo{booktitle}{\emph{16th International Conference
  on World Wide Web, {WWW} 2007}}. \bibinfo{publisher}{{ACM}},
  \bibinfo{address}{Banff, Alberta, Canada}, \bibinfo{pages}{521--530}.
\newblock


\bibitem[\protect\citeauthoryear{Savarese, Silva, and Maire}{Savarese
  et~al\mbox{.}}{2020}]%
        {Cont_Spar}
\bibfield{author}{\bibinfo{person}{Pedro Savarese}, \bibinfo{person}{Hugo
  Silva}, {and} \bibinfo{person}{Michael Maire}.}
  \bibinfo{year}{2020}\natexlab{}.
\newblock \showarticletitle{Winning the Lottery with Continuous
  Sparsification}. In \bibinfo{booktitle}{\emph{Advances in Neural Information
  Processing Systems 33: Annual Conference on Neural Information Processing
  Systems 2020, NeurIPS 2020}}. \bibinfo{publisher}{Curran Associates},
  \bibinfo{address}{virtual}.
\newblock


\bibitem[\protect\citeauthoryear{Schein, Popescul, Ungar, and Pennock}{Schein
  et~al\mbox{.}}{2002}]%
        {Cold-start}
\bibfield{author}{\bibinfo{person}{Andrew~I. Schein},
  \bibinfo{person}{Alexandrin Popescul}, \bibinfo{person}{Lyle~H. Ungar}, {and}
  \bibinfo{person}{David~M. Pennock}.} \bibinfo{year}{2002}\natexlab{}.
\newblock \showarticletitle{Methods and metrics for cold-start
  recommendations}. In \bibinfo{booktitle}{\emph{{SIGIR} 2002: the 25th Annual
  International Conference on Research and Development in Information
  Retrieval}}. \bibinfo{publisher}{{ACM}}, \bibinfo{address}{Tampere, Finland},
  \bibinfo{pages}{253--260}.
\newblock
\urldef\tempurl%
\url{https://doi.org/10.1145/564376.564421}
\showDOI{\tempurl}


\bibitem[\protect\citeauthoryear{Tibshirani}{Tibshirani}{1996}]%
        {LASSO}
\bibfield{author}{\bibinfo{person}{Robert Tibshirani}.}
  \bibinfo{year}{1996}\natexlab{}.
\newblock \showarticletitle{Regression shrinkage and selection via the lasso}.
\newblock \bibinfo{journal}{\emph{Journal of the Royal Statistical Society:
  Series B (Methodological)}} \bibinfo{volume}{58}, \bibinfo{number}{1}
  (\bibinfo{year}{1996}), \bibinfo{pages}{267--288}.
\newblock


\bibitem[\protect\citeauthoryear{Wang, Fu, Fu, and Wang}{Wang
  et~al\mbox{.}}{2017}]%
        {DCN}
\bibfield{author}{\bibinfo{person}{Ruoxi Wang}, \bibinfo{person}{Bin Fu},
  \bibinfo{person}{Gang Fu}, {and} \bibinfo{person}{Mingliang Wang}.}
  \bibinfo{year}{2017}\natexlab{}.
\newblock \showarticletitle{Deep \& Cross Network for Ad Click Predictions}. In
  \bibinfo{booktitle}{\emph{ADKDD'17}} \emph{(\bibinfo{series}{ADKDD'17})}.
  \bibinfo{publisher}{Association for Computing Machinery},
  \bibinfo{address}{Canada}, Article \bibinfo{articleno}{12},
  \bibinfo{numpages}{7}~pages.
\newblock


\bibitem[\protect\citeauthoryear{Wang, Zhao, Xu, and Wu}{Wang
  et~al\mbox{.}}{2022}]%
        {AutoField}
\bibfield{author}{\bibinfo{person}{Yejing Wang}, \bibinfo{person}{Xiangyu
  Zhao}, \bibinfo{person}{Tong Xu}, {and} \bibinfo{person}{Xian Wu}.}
  \bibinfo{year}{2022}\natexlab{}.
\newblock \showarticletitle{AutoField: Automating Feature Selection in Deep
  Recommender Systems}. In \bibinfo{booktitle}{\emph{{WWW} '22: The {ACM} Web
  Conference 2022}}. \bibinfo{publisher}{{ACM}}, \bibinfo{address}{Virtual
  Event, Lyon, France}, \bibinfo{pages}{1977--1986}.
\newblock
\urldef\tempurl%
\url{https://doi.org/10.1145/3485447.3512071}
\showDOI{\tempurl}


\bibitem[\protect\citeauthoryear{Wu, Dai, Zhang, Wang, Sun, Wu, Tian, Vajda,
  Jia, and Keutzer}{Wu et~al\mbox{.}}{2019}]%
        {fbnet}
\bibfield{author}{\bibinfo{person}{Bichen Wu}, \bibinfo{person}{Xiaoliang Dai},
  \bibinfo{person}{Peizhao Zhang}, \bibinfo{person}{Yanghan Wang},
  \bibinfo{person}{Fei Sun}, \bibinfo{person}{Yiming Wu},
  \bibinfo{person}{Yuandong Tian}, \bibinfo{person}{Peter Vajda},
  \bibinfo{person}{Yangqing Jia}, {and} \bibinfo{person}{Kurt Keutzer}.}
  \bibinfo{year}{2019}\natexlab{}.
\newblock \showarticletitle{FBNet: Hardware-Aware Efficient ConvNet Design via
  Differentiable Neural Architecture Search}. In
  \bibinfo{booktitle}{\emph{{IEEE} Conference on Computer Vision and Pattern
  Recognition, {CVPR} 2019}}. \bibinfo{publisher}{Computer Vision Foundation /
  {IEEE}}, \bibinfo{address}{Long Beach, CA, USA},
  \bibinfo{numpages}{10734--10742}~pages.
\newblock
\urldef\tempurl%
\url{https://doi.org/10.1109/CVPR.2019.01099}
\showDOI{\tempurl}


\bibitem[\protect\citeauthoryear{Xie, Wang, Li, Ding, G{\"{u}}rel, Zhang,
  Huang, Lin, and Zhou}{Xie et~al\mbox{.}}{2021}]%
        {FIVES}
\bibfield{author}{\bibinfo{person}{Yuexiang Xie}, \bibinfo{person}{Zhen Wang},
  \bibinfo{person}{Yaliang Li}, \bibinfo{person}{Bolin Ding},
  \bibinfo{person}{Nezihe~Merve G{\"{u}}rel}, \bibinfo{person}{Ce Zhang},
  \bibinfo{person}{Minlie Huang}, \bibinfo{person}{Wei Lin}, {and}
  \bibinfo{person}{Jingren Zhou}.} \bibinfo{year}{2021}\natexlab{}.
\newblock \showarticletitle{{FIVES:} Feature Interaction Via Edge Search for
  Large-Scale Tabular Data}. In \bibinfo{booktitle}{\emph{{KDD} '21: The 27th
  {ACM} {SIGKDD} Conference on Knowledge Discovery and Data Mining}}.
  \bibinfo{publisher}{{ACM}}, \bibinfo{address}{Virtual Event, Singapore},
  \bibinfo{pages}{3795--3805}.
\newblock
\urldef\tempurl%
\url{https://doi.org/10.1145/3447548.3467066}
\showDOI{\tempurl}


\bibitem[\protect\citeauthoryear{Xie, Zhang, Zhu, Huang, and Lin}{Xie
  et~al\mbox{.}}{2020}]%
        {feature_spar}
\bibfield{author}{\bibinfo{person}{Zhenda Xie}, \bibinfo{person}{Zheng Zhang},
  \bibinfo{person}{Xizhou Zhu}, \bibinfo{person}{Gao Huang}, {and}
  \bibinfo{person}{Stephen Lin}.} \bibinfo{year}{2020}\natexlab{}.
\newblock \showarticletitle{Spatially Adaptive Inference with Stochastic
  Feature Sampling and Interpolation}. In \bibinfo{booktitle}{\emph{Computer
  Vision - {ECCV} 2020 - 16th European Conference}}
  \emph{(\bibinfo{series}{Lecture Notes in Computer Science},
  Vol.~\bibinfo{volume}{12346})}. \bibinfo{publisher}{Springer},
  \bibinfo{address}{Glasgow, UK}, \bibinfo{pages}{531--548}.
\newblock
\urldef\tempurl%
\url{https://doi.org/10.1007/978-3-030-58452-8\_31}
\showDOI{\tempurl}


\bibitem[\protect\citeauthoryear{Yuan, Savarese, and Maire}{Yuan
  et~al\mbox{.}}{2021}]%
        {Grow_Spar}
\bibfield{author}{\bibinfo{person}{Xin Yuan}, \bibinfo{person}{Pedro
  Henrique~Pamplona Savarese}, {and} \bibinfo{person}{Michael Maire}.}
  \bibinfo{year}{2021}\natexlab{}.
\newblock \showarticletitle{Growing Efficient Deep Networks by Structured
  Continuous Sparsification}. In \bibinfo{booktitle}{\emph{9th International
  Conference on Learning Representations, {ICLR} 2021}}.
  \bibinfo{publisher}{OpenReview.net}, \bibinfo{address}{Virtual Event,
  Austria}.
\newblock


\bibitem[\protect\citeauthoryear{Zhang, Sheng, Zhang, Jiang, Han, Deng, and
  Zheng}{Zhang et~al\mbox{.}}{2022}]%
        {CTR-Overfit}
\bibfield{author}{\bibinfo{person}{Zhao{-}Yu Zhang},
  \bibinfo{person}{Xiang{-}Rong Sheng}, \bibinfo{person}{Yujing Zhang},
  \bibinfo{person}{Biye Jiang}, \bibinfo{person}{Shuguang Han},
  \bibinfo{person}{Hongbo Deng}, {and} \bibinfo{person}{Bo Zheng}.}
  \bibinfo{year}{2022}\natexlab{}.
\newblock \showarticletitle{Towards Understanding the Overfitting Phenomenon of
  Deep Click-Through Rate Prediction Models}.
\newblock \bibinfo{journal}{\emph{CoRR}}  \bibinfo{volume}{abs/2209.06053}
  (\bibinfo{year}{2022}).
\newblock
\urldef\tempurl%
\url{https://doi.org/10.48550/arXiv.2209.06053}
\showDOI{\tempurl}
\showeprint[arXiv]{2209.06053}


\bibitem[\protect\citeauthoryear{Zhu, Liu, Yang, Zhang, and He}{Zhu
  et~al\mbox{.}}{2021}]%
        {fuxictr}
\bibfield{author}{\bibinfo{person}{Jieming Zhu}, \bibinfo{person}{Jinyang Liu},
  \bibinfo{person}{Shuai Yang}, \bibinfo{person}{Qi Zhang}, {and}
  \bibinfo{person}{Xiuqiang He}.} \bibinfo{year}{2021}\natexlab{}.
\newblock \showarticletitle{Open Benchmarking for Click-Through Rate
  Prediction}. In \bibinfo{booktitle}{\emph{30th ACM International Conference
  on Information \& Knowledge Management}}. \bibinfo{publisher}{Association for
  Computing Machinery}, \bibinfo{address}{Australia},
  \bibinfo{pages}{2759–2769}.
\newblock


\end{thebibliography}
